\documentclass[prb,twocolumn,amssymb,amsmath,eqsecnum,%
floatfix,showpacs]{revtex4}
\usepackage{bm,dcolumn}
\newcommand{\vect}[1]{\mathbf{#1}}
\newcommand{\order}{O}

\DeclareMathOperator{\Tr}{Tr}
\DeclareMathOperator{\tr}{tr}
\begin{document}
\title{Quadratic response theory for spin-orbit coupling in
       semiconductor heterostructures}
\author{Bradley A. Foreman}
\email{phbaf@ust.hk}
\affiliation{Department of Physics,
             Hong Kong University of Science and Technology,
             Clear Water Bay, Kowloon, Hong~Kong, China}

\begin{abstract}
This paper examines the properties of the self-energy operator in
lattice-matched semiconductor heterostructures, focusing on
nonanalytic behavior at small values of the crystal momentum, which
gives rise to long-range Coulomb potentials.  A nonlinear response
theory is developed for nonlocal spin-dependent perturbing potentials.
The ionic pseudopotential of the heterostructure is treated as a
perturbation of a bulk reference crystal, and the self-energy is
derived to second order in the perturbation.  If spin-orbit coupling
is neglected outside the atomic cores, the problem can be analyzed as
if the perturbation were a local spin scalar, since the nonlocal
spin-dependent part of the pseudopotential merely renormalizes the
results obtained from a local perturbation.  The spin-dependent terms
in the self-energy therefore fall into two classes: short-range
potentials that are analytic in momentum space, and long-range
nonanalytic terms that arise from the screened Coulomb potential
multiplied by a spin-dependent vertex function.  For an insulator at
zero temperature, it is shown that the electronic charge induced by a
given perturbation is exactly linearly proportional to the charge of
the perturbing potential.  These results are used in a subsequent
paper to develop a first-principles effective-mass theory with
generalized Rashba spin-orbit coupling.
\end{abstract}


\maketitle

\section{Introduction}

The Rashba Hamiltonian \cite{Rashba60} is the prototype of a class of
effective-mass Hamiltonians describing spin-orbit coupling in
semiconductors. \cite{Wink03} These models have been under intensive
study in the past several years due to theoretical and experimental
advances in spin-related phenomena such as the intrinsic spin Hall
effect
\cite{MurNagZha03,MurNagZha04a,Sin04,Culcer04,Shen04,Shen05,Ma04,%
MurNagZha04b,SchlieLoss05,BernZhang05a,BernZhang04b,ZhangYang05,Hu05,%
Kato04,WuKaSiJu05} and the spin galvanic and circular photogalvanic
effects.
\cite{Gan01,Gan02a,Gan02b,Golub03,Gan03b,Belkov03,Gan03c,Gan04a} Such
effects are generated by spin-orbit coupling terms in the conduction
or valence bands of clean nonmagnetic semiconductors.  Although a
variety of different two- and three-dimensional semiconductor systems
are under investigation, one of the most widely studied is a
heterostructure between semiconductors with the zinc-blende structure,
in which an external electric field can be used to tune the relative
contributions from the Rashba and Dresselhaus spin-splitting
terms. \cite{Nitta97,Schliemann03,Gan04a}

In a two-dimensional effective-mass model, the Rashba coupling has no
coordinate dependence.  But in a three-dimensional theory, it is
usually separated into (1)~a contribution proportional to the
macroscopic electric field generated by gate voltages, dopants, and
free carriers; and (2)~$\delta$ functions representing the
contribution from the rapid change in potential at a heterojunction.
\cite{Leib77,Vasko79,Bast81,Bast82,Las85,LMR85,GerSub92,Fore93,%
PfZa95,AndRocBas97,SchEngLang98,Wink00,Wink03} However, the assumption
that a heterojunction can be represented by a short-range $\delta$
potential has never been justified from first principles.  In a
self-consistent theory with electron-electron interactions, there are
in general long-range Coulomb multipole potentials that are not well
localized at the interface.  These long-range potentials contribute
spin-dependent terms to the Hamiltonian.  Thus, it is important to
establish the conditions under which such terms will appear in the
effective-mass Hamiltonian for a heterojunction.

This paper examines the long-range terms in the self-energy of a
quasiparticle for the case of a lattice-matched semiconductor
heterostructure.  The potential energy of the ions is described using
norm-conserving pseudopotentials.  \cite{BaHaSc82,GoTeHu96,HaGoHu98}
The heterostructure pseudopotential is treated as a perturbation of a
bulk reference crystal, with the self-energy calculated to second
order in the perturbation using quadratic response
theory. \cite{HuZar88,Pit95,Berg97,Berg99,Naz02,Nag03} This approach
is well justified numerically, since the linear response alone has
been shown to give excellent predictions for the valence band offset
in a variety of material systems, including isovalent and heterovalent
heterostructures. \cite{ResBarBal89,BaReBa88,BaReBaPe89,PBBR90,PBRB91,%
ColResBar91,BaPeReBa92,PerBar94,MonPer96}

The approach used here follows closely earlier work by Sham
\cite{Sham66} on the theory of shallow impurity states in bulk
semiconductors.  Sham's work is generalized to include nonlocal
spin-dependent potentials and terms of higher order in the crystal
momentum.  Even for local spin-independent potentials, the present
work includes terms neglected in Sham's analysis, such as dipole
potentials in the quadratic response.

This paper is limited to a study of the electron self-energy in the
limit of small crystal momentum.  The derivation of an effective-mass
Hamiltonian from these results is presented in the following
paper. \cite{Fore05b}

As a workable approximation scheme, the calculation of the linear
response is carried out to terms two orders in $q$ higher than the
lowest nonvanishing term, where $\vect{q}$ is the crystal momentum
transfer of the perturbing potential.  The quadratic response is
calculated to the same order in $q$ as the lowest nonvanishing term in
the linear response.  (See the following paper \cite{Fore05b} for
further discussion of this approximation scheme.)  Three classes of
heterostructure perturbations are considered:

(I) Heterovalent perturbations with nonzero charge.  In this case the
perturbing potential includes a monopole of $\order (q^{-2})$,
and the analysis is performed to an accuracy of $\order (q^0)$ in
the linear response and $\order (q^{-2})$ in the quadratic
response.

(II) Isovalent perturbations for which the linear response has a
nonzero dipole moment.  In this case the leading term in the linear
response is $\order (q^{-1})$, so the linear response is
evaluated to $\order (q)$ and the quadratic response is evaluated
to $\order (q^{-1})$.

(III) Isovalent perturbations for which the linear response has no
dipole moment.  In this case the leading term in the linear response
is $\order (q^{0})$, so the linear response is evaluated to
$\order (q^2)$ and the quadratic response is evaluated to
$\order (q^{0})$.

The perturbations that make up a given heterostructure are generally a
mixture of classes I, II, and III.  The simplest situation is that of
an isovalent heterostructure made up of semiconductors with the
zinc-blende structure, such as GaAs/AlAs or InAs/GaSb.  In this case,
every ionic perturbation is an isovalent perturbation from class III.

Most theoretical and experimental studies of the Rashba spin-splitting
Hamiltonian have dealt with this type of heterostructure.  This case
is therefore studied in greatest detail here, by working out the
explicit form of the self-energy from crystal symmetry.  The results
show that in this case the Rashba Hamiltonian contains only
short-range terms (to within the accuracy of the stated approximation
scheme).  However, there are long-range spin-dependent terms that are
not of the Rashba form.

In a heterovalent system such as Ge/GaAs, the ionic perturbations are
from class I.  However, since macroscopic accumulations of charge are
energetically unfavorable, real heterostructures tend to be
macroscopically neutral.  \cite{BaReBaPe89,Har78} Such a nominally
heterovalent class I problem can therefore often be reduced to an
isovalent class II or III problem by grouping the ions together in
neutral clusters and treating these clusters as the basic unit.  This
approach is discussed further in Appendix \ref{app:cluster}.

In wurtzite heterostructures such as GaN/AlN, the ionic perturbations
are from class II, since the site symmetry of atoms in the wurtzite
structure (space group $C_{6v}^4$) permits a dipole moment.  These
dipole terms produce spontaneous polarization along the hexagonal $c$
axis in bulk wurtzite crystals, leading to macroscopic interface
charge at heterojunctions.
\cite{VaKiSm93,BernFiorVand97a,BernFiorVand97b,BernFior98,FiorBern99,%
BechGroFur00,BernFiorVand01,Zoroddu01} Such charge produces
macroscopic electric fields that generate different piezoelectric
strain fields in different materials.  The present theory, which is
restricted to lattice-matched heterostructures, is therefore not
generally applicable to wurtzite systems (except in the unrealistic
special case \cite{VaKiSm93} where the interface polarization charge
is exactly cancelled by an external interface charge).  However, the
results derived here provide a first step towards a more general
theory dealing with lattice-mismatched heterostructures.

The paper begins in Sec.\ \ref{sec:green} by establishing the basic
definitions and notation for the Green function and self-energy that
are used throughout the paper.  The finite-temperature formalism is
used (both for generality and because it facilitates the derivation of
Ward identities), although the main interest of this paper is the
limit of an insulator at zero temperature.  In Sec.\ \ref{sec:vertex},
the self-energy is expanded in powers of the perturbing potential
using vertex functions.  A set of Ward identities is derived for the
vertex functions at finite and zero temperature.  Section
\ref{sec:density} presents general expressions for the nonlinear
density response, including Ward identities for the static
polarization.  The bare ionic perturbations are screened in Sec.\
\ref{sec:screening}, where the proper vertex functions and proper
polarization are introduced.

A detailed analysis of the small wave vector properties of the linear
screened potential is carried out in Sec.\ \ref{sec:linear} for the
special case of a local spin-independent perturbation.  The quadratic
response for the same case is considered in Sec.\ \ref{sec:quadratic},
and the linear and quadratic contributions to the self-energy are
derived in Sec.\ \ref{sec:selfenergy}.  In Sec.\ \ref{sec:nonlocal} it
is shown that in the norm-conserving pseudopotential formalism, the
contributions from the nonlocal spin-dependent part of the perturbing
potential merely renormalize the contributions from the local part of
the perturbation.  The main results of the paper are discussed and
summarized in Sec.\ \ref{sec:conclusions}.

\section{Green function and self-energy}

\label{sec:green}

This section establishes the notation, basic definitions, and symmetry
properties of the Green function and self-energy used in subsequent
sections of the paper.  The starting point is the definition of the
one-particle thermal Green function
\cite{AbrGorDzy75,FetWal03,NegOrl98}
\begin{equation}
   G_{ss'}(\vect{x}, \tau; \vect{x}', \tau') = - \langle T_{\tau}
   [\hat{\psi}_s(\vect{x}, \tau) \hat{\psi}_{s'}^{\dag}(\vect{x}',
   \tau') ] \rangle , \label{eq:G_def}
\end{equation}
where $s = \pm \frac12$ labels the $z$ component of the spin, $\tau$
is the imaginary time, $T_{\tau}$ is the time ordering operator, and
$\hat{\psi}_s(\vect{x}, \tau) = e^{\hat{K} \tau}
\hat{\psi}_s(\vect{x}) e^{-\hat{K} \tau}$ and
$\hat{\psi}_s^{\dag}(\vect{x}, \tau) = e^{\hat{K} \tau}
\hat{\psi}_s^{\dag}(\vect{x}) e^{-\hat{K} \tau}$ are field operators
in the Heisenberg picture.  The angular brackets denote a thermal
average
\begin{equation}
   \langle \hat{O} \rangle = e^{\beta \Omega} \Tr (e^{-\beta \hat{K}}
   \hat{O} ) , \label{eq:thermal_ave}
\end{equation}
where $\beta = 1 / k_B T$ is the inverse temperature, $\Tr$ denotes a
trace over the many-particle Fock space, $\hat{K} = \hat{H} - \mu
\hat{N}$ is the grand Hamiltonian (with $\mu$ the chemical potential
and $\hat{N}$ the number operator), and $e^{-\beta \Omega} = \Tr
(e^{-\beta \hat{K}})$.  The many-particle Hamiltonian is defined by
\begin{multline}
   \hat{H} = \sum_{s,s'} \iint \hat{\psi}_s^{\dag}(\vect{x})
   h_{ss'}(\vect{x}, \vect{x}') \hat{\psi}_{s'} (\vect{x}') d^3 \! x
   \, d^3 \! x' \\ + \frac12 \sum_{s,s'} \iint
   \frac{\hat{\psi}_s^{\dag}(\vect{x}) \hat{\psi}_{s'}^{\dag}
   (\vect{x}') \hat{\psi}_{s'} (\vect{x}') \hat{\psi}_s (\vect{x})}{|
   \vect{x} - \vect{x}' |} d^3 \! x \, d^3 \! x' ,
\end{multline}
where $h = h^{\dag}$ is the Hamiltonian of a single noninteracting
particle, and Hartree atomic units are used.

Since $\hat{K}$ is time independent, $G$ has the form
$G_{ss'}(\vect{x}, \tau; \vect{x}', \tau') = G_{ss'}(\vect{x},
\vect{x}', \tau - \tau')$, with $G_{ss'}(\vect{x}, \vect{x}', \tau -
\beta) = -G_{ss'}(\vect{x}, \vect{x}', \tau)$ for $0 < \tau <
\beta$.  This permits the Fourier series representation (for $-\beta <
\tau < \beta$)
\begin{subequations}
\begin{align}
   G(\tau) & = \frac{1}{\beta} \sum_{n = -\infty}^{\infty} G(\zeta_n)
   e^{-i \zeta_n \tau} , \\ G(\zeta_n) & = \int_{0}^{\beta} G(\tau)
   e^{i \zeta_n \tau} d\tau ,
\end{align}
\end{subequations}
where $\zeta_n = (2n + 1) \pi / \beta$, and $G(\tau)$ denotes a
single-particle operator whose matrix elements in the $|\vect{x}, s
\rangle$ basis are $G_{ss'}(\vect{x}, \vect{x'}, \tau)$.  A continuous
Green function $G(\omega)$ may then be defined by analytic
continuation of $G(\zeta_n)$ from the discrete frequencies $\omega =
\mu + i \zeta_n$.

The Green function $G(\omega)$ satisfies Dyson's equation
\begin{equation}
   [\omega - h - \Sigma(\omega)] G(\omega) = G(\omega)
   [\omega - h - \Sigma(\omega)] = 1 , \label{eq:Dyson1}
\end{equation}
which is an implicit definition for the self-energy operator
\begin{equation}
   \Sigma(\omega) = \omega - h - G^{-1}(\omega) .
   \label{eq:Sigma_def}
\end{equation}
A formal solution to Eq.\ (\ref{eq:Dyson1}) can be constructed by
solving the nonhermitian eigenvalue equations
\begin{subequations}
\label{eq:Dyson2}
\begin{align}
   [ h + \Sigma(\omega) ] | \psi_n (\omega) \rangle & =
   E_n(\omega) | \psi_n (\omega) \rangle , \label{eq:Dyson2a} \\ 
   [ h + \Sigma^{\dag}(\omega) ] | \chi_n (\omega) \rangle & =
   E_n^*(\omega) | \chi_n (\omega) \rangle , \label{eq:Dyson2b}
\end{align}
\end{subequations}
which are also referred to as Dyson's equations.  It is usually
assumed that the solutions to (\ref{eq:Dyson2}) form a complete
biorthonormal \cite{MorsFesh53_p884} set with the properties
\begin{subequations}
\label{eq:biorth}
\begin{gather}
   \langle \chi_n (\omega) | \psi_{n'} (\omega) \rangle =
   \delta_{nn'} , \\
   \sum_n | \psi_n (\omega) \rangle \langle \chi_n (\omega) | = 1
   ,
\end{gather}
\end{subequations}
although this is difficult to prove in general. \cite{Fried56_p67} If
Eqs.\ (\ref{eq:biorth}) are valid, then $G(\omega)$ is given by
\cite{Layz63,MorsFesh53_p884}
\begin{equation}
   G(\omega) = \sum_n \frac{| \psi_n (\omega) \rangle \langle
   \chi_n (\omega) |}{\omega - E_n(\omega)}
   , \label{eq:G_eig_exp}
\end{equation}
which satisfies the Dyson equation (\ref{eq:Dyson1}) by construction.

Symmetries of the many-particle Hamiltonian $\hat{H}$ under time
reversal and space group operations imply corresponding symmetries of
the one-particle operators $G$ and $\Sigma$.  Derivations of the most
useful symmetry relations are presented in Appendix
\ref{app:symmetry}.

\section{Vertex functions}

\label{sec:vertex}

In this section, a perturbative approach to the Dyson equation
(\ref{eq:Dyson2a}) is developed by using vertex functions to expand
the self-energy in powers of the perturbing potential.

\subsection{Definitions}

The single-particle Hamiltonian $h$ is chosen here to have the form
\begin{equation}
   h_{ss'} (\vect{x}, \vect{x}') = - \tfrac12 \nabla^2 \delta(\vect{x}
   - \vect{x}') \delta_{ss'} + v_{ss'}^{\text{ext}} (\vect{x}, \vect{x}')
   , \label{eq:h}
\end{equation}
where the fixed external potential $v^{\text{ext}}$ is a
norm-conserving ionic pseudopotential,
\cite{BaHaSc82,GoTeHu96,HaGoHu98} which accounts for both spin-orbit
coupling \cite{HybLou86b,Surh91,HeFoNe93,Theur01} and scalar
relativistic effects.  The Dyson equation (\ref{eq:Dyson2a}) is
therefore
\begin{multline}
   - \frac12 \nabla^2 \psi_s(\vect{x}, \omega) + \sum_{s'} \int
   V_{ss'} (\vect{x}, \vect{x}', \omega) \psi_{s'} (\vect{x}', \omega)
   d^3 \! x' \\ = E(\omega) \psi_s (\vect{x}, \omega) ,
   \label{eq:schr_x}
\end{multline}
in which $\psi$ is a spinor wave function, and the total potential
energy $V$ is
\begin{equation}
   V_{ss'} (\vect{x}, \vect{x}', \omega) = v_{ss'}^{\text{ext}}
   (\vect{x}, \vect{x}') + \Sigma_{ss'} (\vect{x}, \vect{x}', \omega)
   . \label{eq:Vtotal}
\end{equation}

The external pseudopotential can be separated as $v^{\text{ext}} =
v^{(0)} + v$, where $v^{(0)}$ is the potential of some periodic
reference crystal, and $v$ is a nonperiodic perturbation associated
with a heterostructure or an impurity.  It is assumed that the total
potential (\ref{eq:Vtotal}) can be represented as a power series in
the perturbation $v$:
\begin{equation}
   V_{s_1 s_2} (\vect{x}_1, \tau_1; \vect{x}_2, \tau_2) \equiv V(12) =
   \sum_{\nu = 0}^{\infty} V^{(\nu)} (12) ,  \label{eq:V_exp}
\end{equation}
where $V^{(0)}$ is the potential (\ref{eq:Vtotal}) when
$v^{\text{ext}} = v^{(0)}$, and the numerical arguments on the
right-hand side are shorthand for the space, spin, and time
coordinates $(1) = (\vect{x}_1, s_1, \tau_1)$.  Although the upper
limit of the formal expansion (\ref{eq:V_exp}) is written as $\nu =
\infty$, this may well be an asymptotic series, and in practice only a
finite number of terms are retained.

The linear and quadratic terms of (\ref{eq:V_exp}) are
\begin{equation} \label{eq:V1V2}
   \begin{split}
   V^{(1)}(12) & = \Gamma^{(1)} (1243) v(34) , \\
   V^{(2)}(12) & = \frac12 \Gamma^{(2)} (124365) v(34) v(56) ,
   \end{split}
\end{equation}
where $\Gamma^{(\nu)}$ is called the vertex function of order $\nu$.
Here a summation or integration of repeated coordinates is assumed,
and the labels are ordered as the trace of a matrix product.  The
perturbation $v$ is taken to be an instantaneous static potential of
the form
\begin{equation}
   v(34) = v_{s_3 s_4} (\vect{x}_3, \vect{x}_4) \delta (\tau_3 -
   \tau_4) . \label{eq:v_static}
\end{equation}
The vertex functions are by definition functional derivatives of $V$
with respect to $v$:
\begin{equation} \label{eq:Ward0}
   \begin{split}
   \Gamma^{(1)}(1243) & = \frac{\delta V(12)}{\delta v(34)} , \\
   \Gamma^{(2)}(124365) & = \frac{\delta^2 V(12)}{\delta v(34) 
   \delta v(56)} \\ & = \frac{\delta \Gamma^{(1)}(1243)}{\delta v(56)} ,
    \end{split}
\end{equation}
which may also be expressed as
\begin{equation} \label{eq:Ward}
   \begin{split} \Gamma^{(1)}(1243) & = \delta(13) \delta(24) +
   \frac{\delta \Sigma(12)}{\delta v(34)} , \\ \Gamma^{(2)}(124365) &
   = \frac{\delta^2 \Sigma(12)}{\delta v(34) \delta v(56)} ,
   \end{split}
\end{equation}
in which $\delta (12) = \delta_{s_1 s_2} \delta(\vect{x}_1 -
\vect{x}_2) \delta (\tau_1 - \tau_2)$.  Note that upon application of
the Fourier transforms defined in Appendix \ref{app:Fourier}, the
above equations hold equally well in momentum and frequency space.

It is convenient at this point to carry out the time integrals in Eq.\
(\ref{eq:V1V2}).  This eliminates the variables $\tau_3, \tau_4,
\ldots$ from $\Gamma^{(\nu)}$ and $v$, and reduces the time
dependence of the $\delta(13) \delta(24)$ term in Eq.\ (\ref{eq:Ward})
to $\delta(\tau_1 - \tau_2)$.  It is assumed below that this has been
done.

\subsection{Ward identities}

\label{subsec:Ward}

The vertex functions satisfy various Ward iden\-ti\-ties
\cite{Noz64,ShamKohn66,JonMar73a} for certain limiting values of their
arguments.  One set of these can be derived by varying the chemical
potential $\mu$ by a small amount $\delta \mu$.  Since $\hat{K} =
\hat{H} - \mu \hat{N}$, this is equivalent to varying $v$ by $\delta
v(12) = -\delta \mu \, \delta (12)$.  For this special case, Eqs.\
(\ref{eq:Ward0}) and (\ref{eq:Ward}) reduce to
\begin{equation}
   \begin{split}
   \Gamma^{(1)}(1233) & = \delta(12) - \frac{\delta
   \Sigma(12)}{\delta \mu} , \\ 
   \Gamma^{(2)}(124355) & = - \frac{\delta \Gamma^{(1)}
   (1243)}{\delta \mu} , 
   \end{split}
\end{equation}
which involve a trace over the input variables of $\Gamma$.  If
$\Sigma$ is analytically continued as a function of $\omega = \mu + i
\zeta_n$, the variation with respect to $\mu$ may be written as
\begin{equation}
   \frac{\delta}{\delta \mu} \rightarrow \frac{\partial}{\partial
   \omega} + \frac{\partial}{\partial \mu} .
\end{equation}

Now for the special case of an insulator at $T = 0$, the chemical
potential can have any value in the range $\mu_{N-1} < \mu < \mu_{N}$,
where $\mu_{N}$ is the minimum energy needed to add one particle to
the ground state of an $N$-particle system, and the energy gap is
$E_{\text{g}} = \mu_{N} - \mu_{N-1}$.  [For small temperatures $\mu$
approaches the well-defined limit $\frac12(\mu_{N} + \mu_{N-1})$,
\cite{AshMer76_p575} but at exactly $T = 0$ it becomes ill defined.]
Since $\mu$ can vary arbitrarily within the gap for a system with
finite $E_{\text{g}}$, the Ward identities for the insulator reduce to
\begin{equation} \label{eq:Ward_Gamma}
   \begin{split} \Gamma^{(1)}(1233) & = \delta(12) - \frac{\partial
   \Sigma(12)}{\partial \omega} , \\ 
   \Gamma^{(2)}(124355) & = - \frac{\partial \Gamma^{(1)}
   (1243)}{\partial \omega} , \\
   \Gamma^{(2)}(123344) & = \frac{\partial^2 \Sigma(12)}{\partial
   \omega^2} ,
   \end{split}
\end{equation}
which generalize and extend the results derived for spin-independent
local potentials in Refs.\ \onlinecite{Sham66} and
\onlinecite{ShamKohn66}.

\section{Nonlinear density response}

\label{sec:density}

\subsection{Definitions}

In this section, perturbation theory (see Appendix
\ref{app:perturbation}) is used to evaluate the electron density of
the system with Hamiltonian $\hat{H} = \hat{H}_0 + \hat{H}_1$, in
which $\hat{H}_0$ is the Hamiltonian of the reference crystal and
$\hat{H}_1$ is the perturbation due to $v$:
\begin{equation}
   \hat{H}_1 = \tr (\hat{\rho} v) = \sum_{s,s'} \iint
   \hat{\rho}_{s's}(\vect{x}', \vect{x}) v_{ss'}(\vect{x}, \vect{x}')
   d^3 \! x \, d^3 \! x' ,
\end{equation}
where $\hat{\rho}$ is the density operator
\begin{equation}
   \hat{\rho}_{s's}(\vect{x}', \vect{x}) =
   \hat{\psi}_s^{\dag}(\vect{x}) \hat{\psi}_{s'} (\vect{x}') .
\end{equation}
The mean nonlocal electron density in the perturbed system is defined
as
\begin{equation}
   n_{ss'}(\vect{x}, \vect{x}') = \langle \hat{\rho}_{ss'}(\vect{x},
   \vect{x}', \tau) \rangle , \label{eq:dens}
\end{equation}
which is independent of $\tau$.  If $n$ is evaluated using the
perturbation theory formula (\ref{eq:pert}), one obtains a power
series in $v$:
\begin{equation}
   n_{ss'}(\vect{x}, \vect{x}') = \sum_{\nu = 0}^{\infty}
   n_{ss'}^{(\nu)}(\vect{x}, \vect{x}') ,
\end{equation}
in which
\begin{equation}
   n_{ss'}^{(0)} (\vect{x}, \vect{x}') = \langle
   \hat{\rho}_{ss'}(\vect{x}, \vect{x}', \tau) \rangle_0 
\end{equation}
is the density of the reference crystal.  (The notation $\langle
\hat{O} \rangle_0$ refers to a thermal average with respect to the
reference crystal; see Appendix \ref{app:perturbation}.)  The terms of
order $\nu > 0$ are given by
\begin{equation}
   n^{(\nu)} (00') = \frac{1}{\nu !} \Pi^{(\nu)} (00', 1'1, \ldots,
   \nu'\nu) v(11') \cdots v(\nu\nu') , \label{eq:n_nu}
\end{equation}
where $\Pi^{(\nu)}$ is the $\nu^{\text{th}}$-order static polarization
(or density correlation function), which is defined in Eq.\
(\ref{eq:Pi_def}).  Here and below, the numerical arguments of
$n^{(\nu)}$, $\Pi^{(\nu)}$, and $v$ are time-independent quantities of
the form $(0) = (\vect{x}_0, s_0)$.

\subsection{Ward identities}

The linear and quadratic density response are given explicitly by
\begin{equation} \label{eq:n1n2}
   \begin{split}
   n^{(1)}(12) & = \Pi^{(1)} (1243) v(34) , \\
   n^{(2)}(12) & = \frac12 \Pi^{(2)} (124365) v(34) v(56) ,
   \end{split}
\end{equation}
which shows that $\Pi$ may be defined as a functional derivative of
$n$ with respect to $v$:
\begin{equation}
   \begin{split}
   \Pi^{(1)}(1243) & = \frac{\delta n(12)}{\delta v(34)} , \\
   \Pi^{(2)}(124365) & = \frac{\delta^2 n(12)}{\delta v(34) 
   \delta v(56)} \\ & = \frac{\delta \Pi^{(1)}(1243)}{\delta v(56)} .
    \end{split}
\end{equation}
For the special case $\delta v(12) = -\delta \mu \, \delta (12)$
representing a variation in chemical potential of $\delta \mu$ (see
Sec.\ \ref{subsec:Ward}), these expressions give
\begin{equation}
   \begin{split}
   \Pi^{(1)}(1233) & = - \frac{\partial n(12)}{\partial
   \mu} , \\
   \Pi^{(2)}(124355) & = - \frac{\partial \Pi^{(1)}
   (1243)}{\partial \mu} ,
   \end{split}
\end{equation}
which are the Ward identities for the static polarization.  For an
insulator at $T = 0$, $\mu$ is indefinite, and these reduce to
\begin{equation}
   \Pi^{(1)}(1233) = 0 , \qquad \Pi^{(2)}(124355) = 0
   . \label{eq:Ward_Pi}
\end{equation}

\section{Screening}

\label{sec:screening}

\subsection{Potential}

In this section, the concept of screening is used to extract the
long-range Coulomb interaction terms from the polarization and vertex
functions.  The first-order screened potential $\varphi$ is defined by
adding the Coulomb potential generated by $n^{(1)}$ to $v$:
\begin{equation}
   \varphi (12) = v(12) + u(1243) \Pi^{(1)} (3465) v(56)
   . \label{eq:phi_0}
\end{equation}
Here $u$ represents the Coulomb interaction, which is spin-independent
and local in coordinate space at both the input and output:
\begin{equation}
   u(1243) = \delta_{s_1 s_2} \delta_{s_3 s_4} \delta(\vect{x}_1 -
   \vect{x}_2) \delta (\vect{x}_3 - \vect{x}_4) u (\vect{x}_1 -
   \vect{x}_3) .
\end{equation}
In momentum space (see Appendix \ref{app:Fourier}) this has the form
\begin{equation}
   u(1243) = \delta_{s_1 s_2} \delta_{s_3 s_4} \delta_{\vect{k}_1 -
   \vect{k}_2, \vect{k}_3 - \vect{k}_4} u (\vect{k}_1 -
   \vect{k}_2) ,
\end{equation}
in which $u(\vect{k}) = v_c(\vect{k}) / \Omega$, where $\Omega$ is the
crystal volume and
\begin{equation}
   v_c(\vect{k}) = \begin{cases}
      4 \pi / k^2  & \text{if $k \ne 0$,} \\
      0 & \text{if $k = 0$.}
   \end{cases}
   \label{eq:vc}
\end{equation}

Another way of writing the screened potential is
\begin{equation}
   \varphi (12) = \epsilon^{-1} (1243) v(34) , \label{eq:phi_1}
\end{equation}
in which the inverse static electronic dielectric matrix is
\begin{equation}
   \epsilon^{-1} (1243) = \delta(13) \delta(24) + u(1265) \Pi^{(1)}
   (5643) . \label{eq:eps_1}
\end{equation}
The dielectric matrix $\epsilon$ satisfies
\begin{equation}
   \epsilon (1243) \epsilon^{-1} (3465) = \epsilon^{-1} (1243)
   \epsilon (3465) = \delta (15) \delta(26) ,
\end{equation}
and is given explicitly below in Eq.\ (\ref{eq:eps}).  For an
insulator, the Ward identity (\ref{eq:Ward_Pi}) yields
\begin{equation}
   \epsilon^{-1} (1233) = \delta(12) . \label{eq:Ward_eps}
\end{equation}
The second-order potential $\varphi^{(2)}$ is just the
Coulomb potential generated by $n^{(2)}$:
\begin{equation}
   \varphi^{(2)} (12) = u(1243) n^{(2)} (34) . \label{eq:phi_2}
\end{equation}

By translation symmetry, $\Pi^{(1)} (1243) = 0$ unless $\vect{k}_1 -
\vect{k}_2 = \vect{k}_3 - \vect{k}_4 + \vect{G}$, where $\vect{G}$ is
a reciprocal lattice vector of the reference crystal.  Equation
(\ref{eq:phi_0}) may therefore be written as
\begin{multline}
   \varphi_{ss'} (\vect{k}, \vect{k}') = v_{ss'} (\vect{k}, \vect{k}')
   + \delta_{ss'} v_c(\vect{q}) \sum_{\vect{k}''} \sum_{\vect{G}} \\
   \times \Pi_{\lambda' \lambda} (\vect{q} ; \vect{k}'', \vect{k}'' +
   \vect{q} + \vect{G}) v_{\lambda \lambda'} (\vect{k}'' + \vect{q} +
   \vect{G}, \vect{k}'') , \label{eq:phi_1a}
\end{multline}
where $\vect{q} \equiv \vect{k} - \vect{k}'$ and
\begin{multline}
   \Pi_{\lambda' \lambda} (\vect{q}; \vect{k}, \vect{k} + \vect{q} +
   \vect{G}) \\ = \frac{1}{\Omega} \sum_{\vect{k}_2} \Pi^{(1)}_{\alpha
   \alpha, \lambda' \lambda} (\vect{q} + \vect{k}_2, \vect{k}_2;
   \vect{k}, \vect{k} + \vect{q} + \vect{G}) . \label{eq:Pi_red}
\end{multline}
This simplified form of $\Pi$ is introduced because the Coulomb
potential depends only on the local spin-independent density
$n(\vect{x}) \equiv n_{\alpha \alpha} (\vect{x}, \vect{x})$.

\subsection{Vertex functions}

Given Eqs.\ (\ref{eq:phi_1}) and (\ref{eq:phi_2}), one can rewrite the
total potentials (\ref{eq:V1V2}) as
\begin{multline} \label{eq:V1V2_phi}
   \begin{aligned}
   V^{(1)}(12) & = \tilde{\Gamma}^{(1)} (1243) \varphi (34) , \\
   V^{(2)}(12) & = \frac12 \tilde{\Gamma}^{(2)} (124365) \varphi(34)
   \varphi(56)
   \end{aligned} 
   \\ + \tilde{\Gamma}^{(1)} (1243) \varphi^{(2)} (34) ,
\end{multline}
in which the {\em proper} vertex function $\tilde{\Gamma}^{(\nu)}$ is
defined in perturbation theory as the sum of all
$\nu^{\text{th}}$-order vertex diagrams that cannot be separated into
two disconnected parts by cutting one Coulomb interaction line or one
electron propagator.  (An alternative definition would be as the
$\nu^{\text{th}}$ functional derivative of $V$ with respect to
$\varphi$.\cite{Hedin65,HedLund69}) From the above results, $\Gamma$
and $\tilde{\Gamma}$ are related by
\begin{multline}
   \begin{aligned}
   \Gamma^{(1)} (1243) & = \tilde{\Gamma}^{(1)} (1265) \epsilon^{-1}
   (5643) , \\
   \Gamma^{(2)} (124365) & = \tilde{\Gamma}^{(2)} (128709)
   \epsilon^{-1} (7843) \epsilon^{-1} (9065) 
   \end{aligned}
   \\ + \tilde{\Gamma}^{(1)} (1287) u(8709) \Pi^{(2)} (904365) .
\end{multline}
In an insulator, the Ward identities (\ref{eq:Ward_Pi}) and
(\ref{eq:Ward_eps}) yield
\begin{equation}
   \begin{split}
   \Gamma^{(1)} (1233) & = \tilde{\Gamma}^{(1)} (1244) , \\
   \Gamma^{(2)} (124355) & = \tilde{\Gamma}^{(2)} (128799)
   \epsilon^{-1} (7843) , \\
   \Gamma^{(2)} (123344) & = \tilde{\Gamma}^{(2)} (125566) .
   \end{split}
\end{equation}
Hence, for insulators, the Ward identities (\ref{eq:Ward_Gamma}) are
valid for both $\Gamma$ and $\tilde{\Gamma}$.

\subsection{Polarization}

\label{sec:screen_polarization}

In a similar fashion, one can define the proper polarization
$\tilde{\Pi}$ as the sum of all static polarization diagrams that
cannot be split by cutting a Coulomb line.  Thus
\begin{equation}
   \Pi^{(1)} (1243) = \tilde{\Pi}^{(1)} (1265) \epsilon^{-1} (5643) ,
   \label{eq:Pi_scr_1}
\end{equation}
which has the form of a Dyson equation: \cite{JonMar73a}
\begin{equation}  \label{eq:Dys_Pi0}
\begin{split}
   \Pi^{(1)} (1243) & = \tilde{\Pi}^{(1)} (1243) +
   \tilde{\Pi}^{(1)}(1265) u(5687) \Pi^{(1)} (7843) \\ & =
   \tilde{\Pi}^{(1)} (1243) + \Pi^{(1)} (1265) u(5687)
   \tilde{\Pi}^{(1)} (7843) .
\end{split}
\end{equation}
This can be used to verify that the dielectric matrix
\begin{equation}
   \epsilon (1243) = \delta(13) \delta(24) - u(1265) \tilde{\Pi}^{(1)}
   (5643)  \label{eq:eps}
\end{equation}
is indeed the inverse of Eq.\ (\ref{eq:eps_1}).  Equations
(\ref{eq:Pi_scr_1}) and (\ref{eq:Dys_Pi0}) can also be written as
\begin{equation}
   \Pi^{(1)} (1243) = \epsilon^{-1} (6512) \tilde{\Pi}^{(1)} (5643) ,
\end{equation}
in which the symmetry property (\ref{eq:D_symm}) was used.  The total
and proper quadratic polarizations are likewise related by (see Fig.\
5 of Ref.\ \onlinecite{Sham66})
\begin{multline}
   \Pi^{(2)} (124365) = \epsilon^{-1} (8712) \tilde{\Pi}^{(2)}
   (78092'1') \epsilon^{-1} (9043) \\ \times \epsilon^{-1} (1'2'65)
   . \label{eq:Pi_scr_2}
\end{multline}

In an insulator, Eqs.\ (\ref{eq:Ward_eps}) and (\ref{eq:Pi_scr_1})
give $\tilde{\Pi}^{(1)} (1233) = \Pi^{(1)} (1244) = 0$, which implies
that $\epsilon (1233) = \delta (12)$.  The inverses of Eqs.\
(\ref{eq:Pi_scr_1}) and (\ref{eq:Pi_scr_2}), i.e.,
\begin{equation}
   \tilde{\Pi}^{(1)} (1243) = \Pi^{(1)} (1265) \epsilon (5643) ,
   \label{eq:Pi_scr_1a}
\end{equation}
\begin{equation}
   \tilde{\Pi}^{(2)} (124365) = \epsilon (8712) \Pi^{(2)} (78092'1')
   \epsilon (9043) \epsilon (1'2'65) , \label{eq:Pi_scr_2a}
\end{equation}
then show that the insulator Ward identities (\ref{eq:Ward_Pi}) are
valid for both $\Pi$ and $\tilde{\Pi}$.

Since the Coulomb interaction depends only on the reduced polarization
matrix (\ref{eq:Pi_red}), Eq.\ (\ref{eq:Dys_Pi0}) can be reduced to
\begin{multline}
   \Pi_{ss'} (\vect{q} + \vect{G}; \vect{k}, \vect{k} + \vect{q} +
   \vect{G}') = \tilde{\Pi}_{ss'} (\vect{q} + \vect{G}; \vect{k},
   \vect{k} + \vect{q} + \vect{G}') \\ + \sum_{\vect{G}''} \tilde{\Pi}
   (\vect{q} + \vect{G}, \vect{q} + \vect{G}'') v_c (\vect{q} +
   \vect{G}'') \\ \times \Pi_{ss'} (\vect{q} + \vect{G}''; \vect{k},
   \vect{k} + \vect{q} + \vect{G}') , \label{eq:Dys_Pi}
\end{multline}
in which a scalar version of $\Pi$ is defined by
\begin{equation}
   \Pi (\vect{q}, \vect{q} + \vect{G}) = \sum_{\vect{k}} \Pi_{\lambda
   \lambda} (\vect{q}; \vect{k}, \vect{k} + \vect{q} + \vect{G})
   . \label{eq:Pi_scalar}
\end{equation}
Now $v_c (\vect{q} + \vect{G}'')$ is nonsingular in the limit $ q
\rightarrow 0$ when $\vect{G}'' \ne \vect{0}$, so it is convenient to
regroup the series expansion of Eq.\ (\ref{eq:Dys_Pi}) so as to
isolate the terms $v_c(\vect{q})$: \cite{AmKo60,JonMar73a_p210}
\begin{multline}
   \Pi_{ss'} (\vect{q} + \vect{G}; \vect{k}, \vect{k} + \vect{q} +
   \vect{G}') = P_{ss'} (\vect{q} + \vect{G}; \vect{k}, \vect{k} +
   \vect{q} + \vect{G}') \\ + P (\vect{q} + \vect{G}, \vect{q}) v_c
   (\vect{q}) \Pi_{ss'} (\vect{q}; \vect{k}, \vect{k} + \vect{q} +
   \vect{G}') . \label{eq:Dys_Pi_P}
\end{multline}
Here $P$ is the sum of all polarization diagrams that cannot be
separated by cutting a Coulomb line labeled with $\vect{q}$ (although
they may be split by cutting lines labeled $\vect{q} + \vect{G}''$
with $\vect{G}'' \ne \vect{0}$).  This will be called the {\em
regular} polarization; it is related to the proper polarization
$\tilde{\Pi}$ by Eq.\ (\ref{eq:Dys_Pi}) with $\Pi \rightarrow P$ and
$\vect{G}'' \ne \vect{0}$.

Both $\tilde{\Pi}$ and $P$ are well behaved in the limit $q
\rightarrow 0$, but $P$ is more convenient for analysis of the
small-$q$ behavior of $\Pi$ because, unlike the case for
$\tilde{\Pi}$, it does not require the inversion of matrices (see
Ref.\ \onlinecite{Onida02} for further discussion and an alternative
approach).  From the relationship between $P$ and $\tilde{\Pi}$, it is
apparent that the insulator Ward identities (\ref{eq:Ward_Pi}) for
$\Pi$ and $\tilde{\Pi}$ hold for $P$ as well.

\section{Linear response to a local perturbation}

\label{sec:linear}

In this section the properties of the screened potential $\varphi$ are
examined in greater detail for the case of a local perturbing
potential, \cite{Sham66} which by definition has the form
\begin{equation} \label{eq:v_local}
   \begin{split}
   v(\vect{x}, \vect{x}') & = \delta (\vect{x} - \vect{x}') v(\vect{x})
   , \\
   v(\vect{k}, \vect{k}') & = v(\vect{k} - \vect{k}') .
   \end{split}
\end{equation}
Here the spin indices were omitted because a hermitian, time-reversal
invariant, local potential must be a spin scalar (see Sec.\
\ref{sec:nonlocal}).  Contributions from the nonlocal part of the
perturbation are considered in Sec.\ \ref{sec:nonlocal}.

\subsection{Screened potential}

With this simplification, all of the polarization matrices can be
reduced to the scalar form (\ref{eq:Pi_scalar}), and the screened
potential (\ref{eq:phi_1a}) simplifies to \cite{Sham66}
\begin{equation}
   \varphi(\vect{q}) = v(\vect{q}) + v_c(\vect{q}) \sum_{\vect{G}}
   \Pi(\vect{q}, \vect{q} + \vect{G}) v(\vect{q} + \vect{G})
   . \label{eq:phi_1b}
\end{equation}
Likewise, the local version of Eq.\ (\ref{eq:Dys_Pi_P}) is
\begin{multline}
   \Pi(\vect{q} + \vect{G}, \vect{q} + \vect{G}') = P(\vect{q} +
   \vect{G}, \vect{q} + \vect{G}') \\ + P(\vect{q} + \vect{G},
   \vect{q}) v_c(\vect{q}) \Pi(\vect{q}, \vect{q} + \vect{G}') .
\end{multline}
It is convenient at this point to define a macroscopic static
electronic dielectric function \cite{JonMar73a_p282}
\begin{equation} \label{eq:eps_k}
\begin{split}
   \epsilon(\vect{k}) & = 1 - v_c(\vect{k}) P(\vect{k}, \vect{k}) \\ &
   = 1 / [1 + v_c(\vect{k}) \Pi(\vect{k}, \vect{k})] ,
\end{split}
\end{equation}
which may be used to express $\Pi$ as a function of the regular
polarization $P$:
\begin{subequations} \label{eq:Pi_P}
\begin{align}
   \Pi(\vect{q}, \vect{q} + \vect{G}) & =
   \epsilon^{-1}(\vect{q}) P(\vect{q}, \vect{q} + \vect{G}) , \\
   \Pi(\vect{q} + \vect{G}, \vect{q}) & = \epsilon^{-1}(\vect{q})
   P(\vect{q} + \vect{G}, \vect{q}) , \\ \Pi(\vect{q} + \vect{G},
   \vect{q} + \vect{G}') & = P(\vect{q} + \vect{G}, \vect{q} +
   \vect{G}') + P(\vect{q} + \vect{G}, \vect{q}) \nonumber \\ & \times
   \epsilon^{-1}(\vect{q}) v_c(\vect{q}) P(\vect{q}, \vect{q} +
   \vect{G}') .
\end{align}
\end{subequations}
Here all of the nonanalytic behavior at small $q$ is contained in the
factors $v_c(\vect{q})$ and $\epsilon^{-1}(\vect{q})$.

Upon substituting (\ref{eq:Pi_P}) into (\ref{eq:phi_1b}), one obtains
the screened potentials
\begin{equation}
   \varphi(\vect{q}) = \frac{v(\vect{q})}{\epsilon(\vect{q})} +
   \frac{v_c(\vect{q})}{\epsilon(\vect{q})} \sum_{\vect{G} \ne
   \vect{0}} P(\vect{q}, \vect{q} + \vect{G}) v(\vect{q} + \vect{G}) ,
   \label{eq:phi_q}
\end{equation}
\begin{multline}
   \varphi(\vect{q} + \vect{G}) = v(\vect{q} + \vect{G}) +
   v_c(\vect{q} + \vect{G}) P(\vect{q} + \vect{G}, \vect{q})
   \varphi(\vect{q}) \\ + v_c(\vect{q} + \vect{G}) \sum_{\vect{G}' \ne
   \vect{0}} P(\vect{q} + \vect{G}, \vect{q} + \vect{G}') v(\vect{q} +
   \vect{G}') , \label{eq:phi_qG}
\end{multline}
where both expressions are valid for arbitrary $\vect{q}$ and
$\vect{G}$, but the latter is more useful for investigating the
behavior of $\varphi$ in the neighborhood of a nonzero reciprocal
lattice vector.  For small $\vect{q}$, the first term in
(\ref{eq:phi_q}) is the macroscopic screening that occurs even for
slowly varying potentials [with $v(\vect{k}) = 0$ for $\vect{k}$
outside the first Brillouin zone], while the second term is a
local-field correction \cite{Louie75,Ortiz90} arising from the
microscopic inhomogeneity of the reference crystal.

\subsection{Power series expansions}

The next step is to establish the small-$q$ properties of $P$.  Since
the only singular Coulomb terms in $P(\vect{q} + \vect{G}, \vect{q} +
\vect{G}')$ are the factors $v_c(\vect{q} + \vect{G}'')$ with
$\vect{G}'' \ne \vect{0}$, the regular polarization $P(\vect{q} +
\vect{G}, \vect{q} + \vect{G}')$ is analytic for $q < G_{\text{min}}$,
where $G_{\text{min}}$ is the magnitude of the smallest nonzero
reciprocal lattice vector.  In addition, the symmetry property
(\ref{eq:D_symm}) gives $P(\vect{x}, \vect{x}') = P(\vect{x}',
\vect{x})$ or
\begin{equation}
   P(\vect{k}, \vect{k}') = P(-\vect{k}', -\vect{k}) ,
   \label{eq:P_sym_k}
\end{equation}
while the Ward identity (\ref{eq:Ward_Pi}) for an insulator implies
that
\begin{equation}
   \lim_{q \rightarrow 0} P(\vect{q}, \vect{q} + \vect{G}) = 0 .
\end{equation}
From these results, we see that the matrix $P_{\vect{G}\vect{G}'}
(\vect{q}) \equiv P(\vect{q} + \vect{G}, \vect{q} + \vect{G}')$ has
the Taylor series expansion
\begin{subequations} \label{eq:P_Taylor}
\begin{align}
   P_{\vect{0}\vect{0}} (\vect{q}) & = P^{[2]}_{\vect{0}\vect{0}} +
   P^{[4]}_{\vect{0}\vect{0}} + P^{[6]}_{\vect{0}\vect{0}} + 
   \order (q^8) ,
   \label{eq:P_00} \\ P_{\vect{0}\vect{G}} (\vect{q}) & =
   P^{[1]}_{\vect{0}\vect{G}} + P^{[2]}_{\vect{0}\vect{G}} +
   P^{[3]}_{\vect{0}\vect{G}} + P^{[4]}_{\vect{0}\vect{G}} +
    \order (q^5) ,
   \label{eq:P_0G} \\ P_{\vect{G}\vect{G}'} (\vect{q}) & =
   P^{[0]}_{\vect{G}\vect{G}'} + P^{[1]}_{\vect{G}\vect{G}'} +
   P^{[2]}_{\vect{G}\vect{G}'} + \order (q^3) , \label{eq:P_GG}
\end{align}
\end{subequations}
with $P_{\vect{G}\vect{0}} (\vect{q}) = P_{\vect{0},-\vect{G}}
(-\vect{q})$.  Here $P^{[l]}_{\vect{G}\vect{G}'}$ denotes a general
polynomial of order $l$ in the Cartesian components of $\vect{q}$; for
example,
\begin{equation}
   P^{[2]}_{\vect{0}\vect{0}} = P^{[2]}_{\vect{0}\vect{0}} (\vect{q})
   = q_{\alpha} q_{\beta} P^{\alpha\beta}_{\vect{0}\vect{0}} ,
\end{equation}
in which $P^{\alpha\beta}_{\vect{0}\vect{0}}$ is a constant, and a sum
over the Cartesian components $\alpha$ and $\beta$ is implicit.

In the limit of small $q$, the dielectric function of an insulator
therefore tends toward a finite but direction-dependent limit:
\begin{equation}
   \lim_{q \rightarrow 0} \epsilon (\vect{q}) \equiv
   \epsilon(\hat{\vect{q}}) = 1 - 4 \pi
   P^{\alpha\beta}_{\vect{0}\vect{0}} \frac{q_{\alpha} q_{\beta}}{q^2}
   ,
\end{equation}
in which $\hat{\vect{q}} = \vect{q} / q$.  This behavior contributes
nonanalytic terms in the small-$q$ expansion of $\epsilon^{-1}
(\vect{q})$:
\begin{multline}
   \frac{1}{\epsilon(\vect{q})} = \frac{1}{\epsilon(\hat{\vect{q}})} [
   1 + w_c(\vect{q}) ( P^{[4]}_{\vect{0}\vect{0}} +
   P^{[6]}_{\vect{0}\vect{0}} ) + w_c^2(\vect{q}) (
   P^{[4]}_{\vect{0}\vect{0}} )^2 ] \\ + \order (q^6) ,
\end{multline}
in which $w_c(\vect{q}) = v_c(\vect{q}) / \epsilon(\hat{\vect{q}})$.

\subsection{Pseudopotential}

To proceed further it is necessary to make some assumptions about the
perturbing pseudopotential $v(\vect{k})$.  This is taken to be a
superposition of spherically symmetric ionic perturbations, each of
which has the Gaussian form used in Refs.\
\onlinecite{BaHaSc82,GoTeHu96,HaGoHu98}.  The pseudopotential can
therefore be written as
\begin{equation}
   v(\vect{k}) = v_{\text{an}} (\vect{k}) + v_c (\vect{k}) \rho
   (\vect{k}) , \label{eq:v_gauss}
\end{equation}
in which $v_{\text{an}} (\vect{k})$ and $\rho (\vect{k})$ are entire
analytic functions of $\vect{k}$, and $\rho (\vect{k})$ represents a
portion of the pseudocharge density of the perturbation [the other
portion being given by $k^2 v_{\text{an}} (\vect{k}) / 4 \pi$].

For a single ion, these functions have the form of a Gaussian times a
polynomial in $k^2$: \cite{GoTeHu96,HaGoHu98}
\begin{equation}
   \begin{split} \rho (\vect{k}) & = -\frac{Z_v}{\Omega} [1 + \tfrac12
   (k r_0)^2 ] e^{-(k r_0)^2 / 2} , \\ v_{\text{an}} (\vect{k}) & =
   \frac{1}{\Omega} [ 2 \pi Z_v r_0^2 + g (k^2 r_0^2) ] e^{-(k r_0)^2
   / 2} , \end{split}
\end{equation}
where $Z_v$ is the charge of the ion, $r_0$ is a core radius
parameter, and $g(x)$ is a cubic polynomial given in Refs.\
\onlinecite{GoTeHu96} and \onlinecite{HaGoHu98}.  Here $\rho
(\vect{k})$ has been defined in such a way that its Taylor series
contains no term proportional to $k^2$:
\begin{equation}
   \rho(\vect{k}) = \rho_0 + \rho_4 k^4 + \rho_6 k^6 + \cdots .
\end{equation}
Therefore, the only term in $v_c (\vect{k}) \rho(\vect{k})$ that does not
vanish in the limit $k \rightarrow 0$ is the divergent term $-4 \pi
Z_v / \Omega k^2$.  This term has been eliminated \cite{YinCoh82a} at
$k = 0$ [by the definition (\ref{eq:vc}) of $v_c (\vect{k})$] because
the ion is assumed to be accompanied by $Z_v$ electrons, so that the
crystal remains neutral after the perturbation.

If the perturbation contains more than one ion (e.g., the quasiatoms
defined in Appendix \ref{app:cluster}), $\rho(\vect{k})$ is just a
general Taylor series
\begin{equation}
   \rho(\vect{k}) = \rho_0 + \rho^{[1]} + \rho^{[2]} + \cdots ,
\end{equation}
although the linear response can always be treated as a superposition
of individual ions.

For $\vect{G} \ne \vect{0}$, $v(\vect{q} + \vect{G})$ is analytic for
$q < G$, with the Taylor series
\begin{equation}
   v(\vect{q} + \vect{G}) = v_{\vect{G}}^{[0]} + v_{\vect{G}}^{[1]} +
   v_{\vect{G}}^{[2]} + v_{\vect{G}}^{[3]} + \order (q^4) ,
   \label{eq:v_qG}
\end{equation}
in which $v_{\vect{G}}^{[0]} = v(\vect{G})$.  A similar expansion is
valid for $v_c(\vect{q} + \vect{G})$.

\subsection{Effective macroscopic density}

It is now convenient to rewrite Eq.\ (\ref{eq:phi_q}) in a form
modeled after the familiar expressions for screening in a homogeneous
system: \cite{note:screening}
\begin{equation}
   \varphi(\vect{q}) = v_{\text{an}} (\vect{q}) + \frac{v_c (\vect{q})
   \bar{n} (\vect{q})}{\epsilon(\vect{q})} , \label{eq:phi_nbar}
\end{equation}
in which
\begin{equation}
   \bar{n}(\vect{q}) = \rho (\vect{q}) + P_{\vect{0}\vect{0}}
   (\vect{q}) v_{\text{an}} (\vect{q}) + \sum_{\vect{G} \ne \vect{0}}
   P_{\vect{0}\vect{G}} (\vect{q}) v(\vect{q} + \vect{G})
   \label{eq:n_bar}
\end{equation}
is an effective macroscopic electron density, which is analytic for $q
< G_{\text{min}}$.  This includes the (partial) bare charge $\rho
(\vect{q})$, the macroscopic charge induced by $v_{\text{an}}
(\vect{q})$, and the local-field corrections.  Note that since
$\varphi (\vect{q}) = v(\vect{q}) + v_c (\vect{q}) n^{(1)}
(\vect{q})$, one can also write $\bar{n} (\vect{q}) = [\rho (\vect{q})
+ n^{(1)} (\vect{q}) ] \epsilon (\vect{q})$.  Also note that
$\bar{n}_0 = \rho_0$, because the second term in (\ref{eq:n_bar}) is
$\order (q^2)$ and the last term is $\order (q)$.  The leading
contributions to $\varphi(\vect{q})$ for small $q$ are therefore
\begin{multline}
   \varphi(\vect{q}) = \rho_0 [w_c (\vect{q}) + w_c^2 (\vect{q}) (
   P^{[4]}_{\vect{0}\vect{0}} + P^{[6]}_{\vect{0}\vect{0}} ) +
   w_c^3(\vect{q}) ( P^{[4]}_{\vect{0}\vect{0}} )^2 ] \\ +
   w_c(\vect{q}) ( \bar{n}^{[1]} + \bar{n}^{[2]} + \bar{n}^{[3]} +
   \bar{n}^{[4]} ) \\ + w_c^2(\vect{q}) P^{[4]}_{\vect{0}\vect{0}} (
   \bar{n}^{[1]} + \bar{n}^{[2]} ) + v_{\text{an}} (\vect{q}) + \order
   (q^3) , \label{eq:phi_q_exp}
\end{multline}
in which $v_{\text{an}} (\vect{q})$ is to be replaced by its Taylor
series expansion.

The first set of terms in (\ref{eq:phi_q_exp}) is proportional to $\rho_0
= -Z_v / \Omega$.  These terms are just the power series expansion for
$\rho_0 v_c (\vect{q}) / \epsilon (\vect{q})$.  Such terms are present in
general, but they vanish for isovalent perturbations (e.g., Al
substituting for Ga in GaAs).

The remaining nonanalytic terms depend on $\bar{n}^{[l]}$, where $l
\ge 1$.  The symmetry of $\bar{n}^{[l]}$ may differ from that of
$\rho^{[l]}$.  For example, the term $w_c(\vect{q}) n^{[1]}
(\vect{q})$ would contribute a dipole field if it were present, but
$\rho^{[1]}$ vanishes for a spherically symmetric atom.  However,
since the symmetry of $P$ is the same as that of the reference
crystal, the symmetry of $\bar{n}$ is just the site symmetry at the
position of the ionic perturbation (i.e., the maximal common subgroup
of the reference crystal space group and the full rotation group at
the given atomic site).  Therefore, for crystals of sufficiently low
symmetry (e.g., wurtzite), $\bar{n}^{[1]}$ may contribute a dipole
field even though $\rho^{[1]}$ does not.

\subsection{Special cases}

The general expression (\ref{eq:phi_q_exp}) is quite cumbersome and is
unlikely to be used in its entirety for any particular material
system.  In many cases one would only be interested in retaining terms
that are two orders in $q$ higher than the lowest nonvanishing term.
Thus, for heterovalent perturbations with $\rho_0 \ne 0$ (i.e., class I
of the Introduction), Eq.\ (\ref{eq:phi_q_exp}) could be simplified to
\begin{multline}
   \varphi(\vect{q}) = v_{\text{an}} (\vect{0}) + w_c(\vect{q}) (
   \rho_0 + \bar{n}^{[1]} + \bar{n}^{[2]} ) \\ + \rho_0 w_c^2
   (\vect{q}) P^{[4]}_{\vect{0}\vect{0}} + \order (q) ,
   \label{eq:phi_q_exp_a}
\end{multline}
which contains monopole, dipole, and quadrupole terms, plus a
correction to the monopole term describing the wave vector dependence
of the dielectric function.  For isovalent perturbations in crystals
with atomic site symmetry that supports a dipole moment (class II), a
suitable approximation would be
\begin{multline}
   \varphi(\vect{q}) = v_{\text{an}} (\vect{q}) + w_c(\vect{q})
   (\bar{n}^{[1]} + \bar{n}^{[2]} + \bar{n}^{[3]}) \\ +
   w_c^2(\vect{q}) P^{[4]}_{\vect{0}\vect{0}} \bar{n}^{[1]} + \order
   (q^2) , \label{eq:phi_q_exp_b}
\end{multline}
which includes additional octopole terms.  Finally, for isovalent
perturbations in crystals with site symmetry that does not support a
dipole moment (class III), one would use
\begin{multline}
   \varphi(\vect{q}) = v_{\text{an}} (\vect{q}) + w_c(\vect{q})
   (\bar{n}^{[2]} + \bar{n}^{[3]} + \bar{n}^{[4]} ) \\ +
   w_c^2(\vect{q}) P^{[4]}_{\vect{0}\vect{0}} \bar{n}^{[2]} + \order
   (q^3) .  \label{eq:phi_q_exp_c}
\end{multline}

As an explicit example, consider the case of isovalent substitutions
in a crystal with the zinc-blende or diamond structure (space group
$T_d^2$ or $O_h^7$), both of which have site symmetry $T_d$ at the
atomic sites.  In this case $\rho_0 = 0$, and the only invariants of
order $q^4$ or lower are 1, $q^2$, $q_x q_y q_z$, $q^4$, and $q_x^4 +
q_y^4 + q_z^4$.  Hence, the quadratic terms are isotropic
($P^{\alpha\beta}_{\vect{0}\vect{0}} = P_2 \delta_{\alpha\beta}$,
$\bar{n}^{[2]} = \bar{n}_2 q^2$) and the long-wavelength dielectric
function reduces to a constant [$\epsilon(\hat{\vect{q}}) = \epsilon =
1 - 4 \pi P_2$].  The leading contributions to $\varphi(\vect{q})$ may
be written as
\begin{multline}
   \varphi(\vect{q}) = \frac{4 \pi \bar{n}_2}{\epsilon} (1 -
   \delta_{\vect{q} \vect{0}}) + C_1 + C_2 q^2 + C_3 \frac{q_x q_y
   q_z}{q^2} \\ + C_4 \frac{q_x^4 + q_y^4 + q_z^4}{q^2} + \order (q^3)
   , \label{eq:multipole1}
\end{multline}
where $C_i$ is a constant.  The terms $C_3$ and $C_4$ represent
octopole and hexadecapole moments, respectively.

\subsection{Nonzero reciprocal lattice vectors}

Turning now to $\varphi(\vect{q} + \vect{G})$, Eq.\ (\ref{eq:phi_qG})
can be written in the condensed notation
\begin{equation}
   \varphi (\vect{q} + \vect{G}) = R_{\vect{G}\vect{0}} (\vect{q})
   \varphi (\vect{q}) + \xi_{\vect{G}} (\vect{q}) ,
   \label{eq:phi_qG_2}
\end{equation}
in which $R_{\vect{G}\vect{0}} (\vect{q})$ and $\xi_{\vect{G}}
(\vect{q})$ are analytic for $q < G_{\text{min}}$:
\begin{equation}
   R_{\vect{G}\vect{0}} (\vect{q}) = \begin{cases}
     1 & \text{if $\vect{G} = \vect{0}$}, \\
     v_c (\vect{q} + \vect{G}) P_{\vect{G}\vect{0}} (\vect{q}) & 
     \text{if $\vect{G} \ne \vect{0}$},
   \end{cases} \label{eq:RG0}
\end{equation}
\begin{multline}
   \xi_{\vect{G}} (\vect{q}) = (1 - \delta_{\vect{G} \vect{0}})
   \biggl[ v(\vect{q} + \vect{G}) + v_c (\vect{q} + \vect{G}) \\
   \times \sum_{\vect{G}' \ne \vect{0}} P_{\vect{G} \vect{G}'}
   (\vect{q}) v (\vect{q} + \vect{G}') \biggr] . \label{eq:chi_G}
\end{multline}
For $\vect{G} \ne \vect{0}$, Eqs.\ (\ref{eq:P_0G}) and (\ref{eq:v_qG})
show that $R_{\vect{G}\vect{0}} (\vect{q})$ is of order $q$ or higher.
The explicit form of the Taylor series for $R_{\vect{G} \vect{0}}
(\vect{q})$ is determined by finding the invariants of the group of
the wave vector $\vect{G}$ in the reference crystal, where different
$\vect{G}$ vectors are treated as inequivalent.  Thus, for general
$\vect{G}$, the linear term is nonvanishing.  However, the leading
term in the Taylor series for $\xi_{\vect{G}} (\vect{q})$ (with
$\vect{G} \ne \vect{0}$) is a constant.

Hence, for $\vect{G} \ne \vect{0}$, the nonanalytic terms in
$\varphi(\vect{q} + \vect{G})$ are at least one order in $q$ higher
than those in $\varphi(\vect{q})$.  In the zinc-blende example
discussed above, one has
\begin{equation}
   \varphi(\vect{q} + \vect{G}) = C_3 R_{\vect{G}\vect{0}}^{\alpha}
   \frac{q_{\alpha} q_x q_y q_z}{q^2} + \chi_{\vect{G}} (\vect{q}) +
   \order (q^3) , \label{eq:multipole2}
\end{equation}
in which $\chi_{\vect{G}} (\vect{q})$ is analytic, and
$R_{\vect{G}\vect{0}}^{\alpha}$ is the linear coefficient in the
Taylor series for $R_{\vect{G}\vect{0}} (\vect{q})$.  The nonanalytic
term in (\ref{eq:multipole2}) is a hexadecapole moment that is
invariant with respect to the group of $\vect{G}$.

\section{Quadratic response to a local perturbation}

\label{sec:quadratic}

To calculate the quadratic density $n^{(2)} (\vect{k})$, it is helpful
to begin by considering the following partial density obtained from
the local version of Eqs.\ (\ref{eq:n1n2}), (\ref{eq:phi_1}), and
(\ref{eq:Pi_scr_2}):
\begin{multline}
   \tilde{n}^{(2)} (\vect{k}) = \frac12 {\sum_{\vect{k}_1}}'
   \sum_{\vect{G}_1 \vect{G}_2} \tilde{\Pi}^{(2)} (\vect{k},
   \vect{k}_1 + \vect{G}_1, \vect{k} - \vect{k}_1 + \vect{G}_2) \\
   \times \varphi (\vect{k}_1 + \vect{G}_1) \varphi (\vect{k} -
   \vect{k}_1 + \vect{G}_2) , \label{eq:n2k}
\end{multline}
where the summation on $\vect{k}_1$ is limited to the first Brillouin
zone of the reference crystal.  Here the proper polarization
$\tilde{\Pi}^{(2)} (\vect{k}, \vect{k}_1, \vect{k}_2)$ vanishes unless
$\vect{k} = \vect{k}_1 + \vect{k}_2 + \vect{G}$, where $\vect{G}$ is
any reciprocal lattice vector.  It satisfies the symmetry relations
(\ref{eq:D_symm}), the reduced form of which is
\begin{equation}
   \tilde{\Pi}^{(2)} (\vect{k}, \vect{k}_1, \vect{k}_2) =
   \tilde{\Pi}^{(2)} (\vect{k}, \vect{k}_2, \vect{k}_1) =
   \tilde{\Pi}^{(2)} (-\vect{k}_1, -\vect{k}, \vect{k}_2) .
\end{equation}
It also satisfies the Ward identity (\ref{eq:Ward_Pi}):
\begin{equation}
   \lim_{k \rightarrow 0} \tilde{\Pi}^{(2)} (\vect{k}, \vect{k}_1,
   \vect{k} - \vect{k}_1 + \vect{G}_2) = 0 . \label{eq:Ward_Pi2}
\end{equation}
With these constraints, the Taylor series expansion of the
polarization matrix $\tilde{\Pi}^{(2)}_{\vect{G} \vect{G}_1
\vect{G}_2} (\vect{k}, \vect{k}_1, \vect{k}_2) = \tilde{\Pi}^{(2)}
(\vect{k} + \vect{G}, \vect{k}_1 + \vect{G}_1, \vect{k}_2 +
\vect{G}_2)$ has a form similar to that given for $P$ in Eq.\
(\ref{eq:P_Taylor}):
\begin{equation} \label{eq:Pi_Taylor}
   \begin{split}
   \tilde{\Pi}^{(2)}_{\vect{0} \vect{0} \vect{0}} (\vect{k},
   \vect{k}_1, \vect{k}_2) & = k_{\alpha} k_{1\beta} k_{2\gamma}
   \tilde{\Pi}^{\alpha\beta\gamma}_{\vect{0} \vect{0} \vect{0}} +
   \order (k^4) , \\
   \tilde{\Pi}^{(2)}_{\vect{G} \vect{0} \vect{0}} (\vect{k},
   \vect{k}_1, \vect{k}_2) & = k_{1\beta} k_{2\gamma}
   \tilde{\Pi}^{\beta\gamma}_{\vect{G} \vect{0} \vect{0}} + 
   \order (k^3) , \\
   \tilde{\Pi}^{(2)}_{\vect{0} \vect{G}_1 \vect{G}_2} (\vect{k},
   \vect{k}_1, \vect{k}_2) & = k_{\alpha}
   \tilde{\Pi}^{\alpha}_{\vect{0} \vect{G}_1 \vect{G}_2} + 
   \order (k^2) , \\
   \tilde{\Pi}^{(2)}_{\vect{G} \vect{G}_1 \vect{G}_2} (\vect{k},
   \vect{k}_1, \vect{k}_2) & = \tilde{\Pi}_{\vect{G}
   \vect{G}_1 \vect{G}_2} + \order (k) ,
   \end{split}
\end{equation}
where the order of the leading term is equal to the number of
$\vect{G}$ vectors that are zero.  Here $\order (k^n)$ denotes a
term of order $k^p k_1^{q} k_2^{r}$, where $n = p + q + r$.

The partial quadratic density (\ref{eq:n2k}) generates a Coulomb
potential $v^{(2)} (\vect{q}) = v_c (\vect{q}) \tilde{n}^{(2)}
(\vect{q})$, which is then screened to produce $\varphi^{(2)}
(\vect{q})$ of Eq.\ (\ref{eq:phi_2}).  This potential is calculated by
replacing $v(\vect{q})$ with $v^{(2)} (\vect{q})$ in Eqs.\
(\ref{eq:phi_q}) and (\ref{eq:phi_qG}).  The result may be written as
\begin{equation}
   \varphi^{(2)} (\vect{q}) = \frac{v_c (\vect{q}) \bar{n}^{(2)}
   (\vect{q})}{\epsilon (\vect{q})} , \label{eq:phi2a}
\end{equation}
where $\bar{n}^{(2)} (\vect{q}) = n^{(2)} (\vect{q}) \epsilon
(\vect{q})$ is an effective ``external'' density
\begin{equation}
   \bar{n}^{(2)} (\vect{q}) = \sum_{\vect{G}} R_{\vect{0} \vect{G}}
   (\vect{q}) \tilde{n}^{(2)} (\vect{q} + \vect{G}) , \label{eq:N2}
\end{equation}
in which
\begin{equation}
   R_{\vect{0} \vect{G}} (\vect{q}) = \begin{cases}
     1 & \text{if $\vect{G} = \vect{0}$}, \\
     P_{\vect{0} \vect{G}} (\vect{q}) v_c (\vect{q} + \vect{G}) & 
     \text{if $\vect{G} \ne \vect{0}$}.
   \end{cases}
\end{equation}
In Eq.\ (\ref{eq:N2}), $\tilde{n}^{(2)} (\vect{q} + \vect{G})$ is
given by (\ref{eq:n2k}), where $\varphi (\vect{k} + \vect{G})$ can be
expressed in terms of $\varphi (\vect{k})$ using Eq.\
(\ref{eq:phi_qG_2}).  The resulting expression for (\ref{eq:N2}) can
be written as $\bar{n}^{(2)} (\vect{q}) = \bar{n}^{(2)}_A (\vect{q}) +
\bar{n}^{(2)}_B (\vect{q}) + \bar{n}^{(2)}_C (\vect{q})$, in which
\begin{equation} \label{eq:N_ABC}
   \begin{split}
   \bar{n}^{(2)}_A (\vect{q}) & = \frac12 {\sum_{\vect{k}}}'
   A(\vect{q}, \vect{k}, \vect{q} - \vect{k}) \varphi (\vect{k})
   \varphi (\vect{q} - \vect{k}) , \\ \bar{n}^{(2)}_B (\vect{q}) & =
   \frac12 {\sum_{\vect{k}}}' [ B(\vect{q}, \vect{k}, \vect{q} -
   \vect{k}) \varphi (\vect{q} - \vect{k}) \\ & \qquad \qquad +
   B(\vect{q}, \vect{q} - \vect{k}, \vect{k}) \varphi (\vect{k}) ] ,
   \\ \bar{n}^{(2)}_C (\vect{q}) & = \frac12 {\sum_{\vect{k}}}'
   C(\vect{q}, \vect{k}, \vect{q} - \vect{k}) .
   \end{split}
\end{equation}
Here the functions $A$, $B$, and $C$, which are defined in Appendix
\ref{app:ABC}, have the Taylor series expansions
\begin{equation} \label{eq:Pi_Taylor_2}
   \begin{split}
   A (\vect{k}, \vect{k}_1, \vect{k}_2) & = k_{\alpha} k_{1\beta}
   k_{2\gamma} A_{\alpha\beta\gamma} + \order (k^4) , \\
   B (\vect{k}, \vect{k}_1, \vect{k}_2) & = k_{\alpha} k_{2\gamma}
   B_{\alpha\gamma} + \order (k^3) , \\
   C (\vect{k}, \vect{k}_1, \vect{k}_2) & = k_{\alpha} C_{\alpha} +
   \order (k^2) .
   \end{split}
\end{equation}

For $\vect{q}$ values inside the first Brillouin zone, the functions
$A$, $B$, and $C$ in (\ref{eq:N_ABC}) are analytic for all $\vect{k}$
values included in the summation, but $\varphi(\vect{k})$ is
nonanalytic at $\vect{k} = \vect{0}$.  Therefore it is possible that
the $\varphi$ terms in (\ref{eq:N_ABC}) may produce nonanalytic
behavior in $\bar{n}^{(2)} (\vect{q})$.  $\bar{n}^{(2)}_C (\vect{q})$
is obviously analytic in $\vect{q}$, as is the second term in
$\bar{n}^{(2)}_B (\vect{q})$.  The first term in $\bar{n}^{(2)}_B
(\vect{q})$ is as well, since a small variation $\delta \vect{q}$ can
be eliminated from $\varphi(\vect{q} - \vect{k})$ with an equal
variation $\delta \vect{k} = \delta \vect{q}$.  This slightly shifts
the zone boundary in the summation, but $\varphi(\vect{q} - \vect{k})$
is analytic at the zone boundary, so $\bar{n}^{(2)}_B (\vect{q})$ is
analytic for small $\vect{q}$.

However, for $\bar{n}^{(2)}_A (\vect{q})$ this argument is no longer
valid.  The singularities in $\varphi(\vect{k})$ and $\varphi(\vect{q}
- \vect{k})$ merge when $\vect{q} = \vect{0}$, producing nonanalytic
behavior in $\bar{n}^{(2)}_A (\vect{q})$ at this point.  The
contribution from the nonanalytic part of $\bar{n}^{(2)}_A (\vect{q})$
is examined in Appendix \ref{app:nonanalytic}, where it is shown to be
negligible under all three approximation schemes defined in the
Introduction.  Therefore, only the analytic part of
$\bar{n}^{(2)}(\vect{q})$ is retained here.

The leading contributions to the quadratic screened potential are
therefore
\begin{equation}
   \varphi^{(2)} (\vect{q}) = w_c (\vect{q}) (q_{\alpha}
   \bar{n}^{(2)}_{\alpha} + q_{\alpha} q_{\beta}
   \bar{n}^{(2)}_{\alpha\beta}) + \order (q) ,
   \label{eq:phi2_q_N}
\end{equation}
where $\bar{n}^{(2)}_{\alpha}$ and $\bar{n}^{(2)}_{\alpha\beta}$ are
Taylor series coefficients for the analytic part of
$\bar{n}^{(2)}(\vect{q})$.  The absence of a constant term in the
power series for $\bar{n}^{(2)}(\vect{q})$ is a consequence of the
Ward identity (\ref{eq:Ward_Pi2}).  Equation (\ref{eq:phi2_q_N}) is
used in its full form only for isovalent class III perturbations.  For
class II, the $\bar{n}^{(2)}_{\alpha\beta}$ term is negligible, while
for class I, the entire contribution from $\varphi^{(2)} (\vect{q})$
is negligible. \cite{note:class1}

In the vicinity of a nonzero reciprocal lattice vector, $\varphi^{(2)}
(\vect{q} + \vect{G})$ can be written in a form similar to
(\ref{eq:phi_qG_2}):
\begin{equation}
   \varphi^{(2)} (\vect{q} + \vect{G}) = R_{\vect{G}\vect{0}}
   (\vect{q}) \varphi^{(2)} (\vect{q}) + \xi^{(2)}_{\vect{G}}
   (\vect{q}) , \label{eq:phi2_qG}
\end{equation}
in which $\xi^{(2)}_{\vect{G}} (\vect{q})$ is given by Eq.\
(\ref{eq:chi_G}) with $v(\vect{k})$ replaced by $v^{(2)} (\vect{k}) =
v_c (\vect{k}) \tilde{n}^{(2)} (\vect{k})$.  Using the same type of
analysis as before, one finds that the nonanalytic part of
$\xi^{(2)}_{\vect{G}} (\vect{q})$ is $\order (q)$ for class I, $\order
(q^3)$ for class II, and $\order (q^5)$ for class III.  Therefore, the
limit $\xi^{(2)}_{\vect{G}} (\vect{0})$ is well defined, and the
leading terms in $\varphi^{(2)} (\vect{q} + \vect{G})$ are given by
\begin{equation}
   \varphi^{(2)} (\vect{q} + \vect{G}) = w_c (\vect{q}) (q_{\alpha}
   q_{\beta} \bar{n}^{(2)}_{\alpha} R_{\vect{G}\vect{0}}^{\beta}) +
   \xi^{(2)}_{\vect{G}} (\vect{0}) + \order (q)
   . \label{eq:phi2_qG_N}
\end{equation}
Since the leading terms here are $\order (q^0)$, this contribution is
negligible for classes I and II.

Note that the quadratic response for a heterostructure cannot be
written as a superposition of spherically symmetric atomic
perturbations; one must also include diatomic perturbations with axial
symmetry $C_{\infty v}$ (for a heteronuclear diatomic molecule) or
$D_{\infty h}$ (for a homonuclear diatomic molecule). \cite{Tink64} The
symmetry of $\bar{n}^{(2)}(\vect{q})$ is determined by the maximal
common subgroup of the reference crystal space group and the molecular
point group.  For example, for a perturbation at neighboring atomic
sites in zinc-blende, the symmetry of $\bar{n}^{(2)}(\vect{q})$ is
$C_{3v}$, which supports a nonvanishing dipole moment
$\bar{n}^{(2)}_{\alpha}$.

In general, $\bar{n}^{(2)}_{\alpha}$ is nonvanishing for any
heteronuclear perturbation, because $C_{\infty v}$ itself permits the
existence of a dipole.  Such dipoles therefore always appear in
heterostructures involving more than one type of atomic perturbation
(e.g., InAs/GaSb).  (This property of the nonlinear response was
deduced from numerical calculations of band offsets in Ref.\
\onlinecite{DanZun92}.)  Furthermore, the quadrupole term
$\bar{n}^{(2)}_{\alpha\beta} - \frac13 \bar{n}^{(2)}_{\lambda\lambda}
\delta_{\alpha\beta}$ is nonvanishing for any diatomic perturbation,
since isotropy requires cubic symmetry.

\begin{widetext}

\section{Self-energy for a local perturbation}

\label{sec:selfenergy}

\subsection{Linear terms}

The above results may now be used to calculate the self-energy
$\Sigma$ and the total potential $V$ defined in Eqs.\
(\ref{eq:Vtotal}) and (\ref{eq:V1V2_phi}).  The total linear potential
(\ref{eq:V1V2_phi}) is given for the case of a local perturbation by
\begin{equation}
   V^{(1)}_{ss'} (\vect{k} + \vect{G}, \vect{k}' + \vect{G}'; \omega)
   = \sum_{\vect{G}''} \tilde{\Gamma}^{(1)}_{ss'} (\vect{k} +
   \vect{G}, \vect{k}' + \vect{G}'; \vect{q} + \vect{G}''; \omega)
   \varphi (\vect{q} + \vect{G}'') , \label{eq:V1_phi}
\end{equation}
where $\vect{q} = \vect{k} - \vect{k}'$.  Here $\varphi (\vect{q} +
\vect{G}'')$ can be expressed in terms of $\varphi (\vect{q})$ using
Eq.\ (\ref{eq:phi_qG_2}); this yields
\begin{equation}
   V^{(1)}_{ss'} (\vect{k} + \vect{G}, \vect{k}' + \vect{G}'; \omega)
   = \Lambda_{ss'} (\vect{k}, \vect{k}'; \vect{G}, \vect{G}'; \omega)
   \varphi (\vect{q}) + W^{(1)}_{ss'} (\vect{k}, \vect{k}'; \vect{G},
   \vect{G}'; \omega) , \label{eq:V1}
\end{equation}
in which
\begin{equation}
   W^{(1)}_{ss'} (\vect{k}, \vect{k}'; \vect{G}, \vect{G}'; \omega) =
   \sum_{\vect{G}'' \ne \vect{0}} \tilde{\Gamma}^{(1)}_{ss'} (\vect{k}
   + \vect{G}, \vect{k}' + \vect{G}'; \vect{q} + \vect{G}''; \omega)
   \xi_{\vect{G}''} (\vect{q}) \label{eq:V1a}
\end{equation}
is an analytic function of $\vect{k}$ and $\vect{k}'$ (and therefore
also of $\vect{q}$).  The nonanalytic terms are all contained in the
screened potential $\varphi(\vect{q}) = v_{\text{an}} (\vect{q}) + v_c
(\vect{q}) \bar{n} (\vect{q}) / \epsilon (\vect{q})$, which is
multiplied by the effective macroscopic vertex function
\begin{equation}
   \Lambda_{ss'} (\vect{k}, \vect{k}'; \vect{G}, \vect{G}'; \omega) =
   \sum_{\vect{G}''} \tilde{\Gamma}^{(1)}_{ss'} (\vect{k} + \vect{G},
   \vect{k}' + \vect{G}'; \vect{q} + \vect{G}''; \omega) R_{\vect{G}''
   \vect{0}} (\vect{q}) . \label{eq:Lambda_ss}
\end{equation}
Since $\Lambda_{ss'}$ is an analytic function of $\vect{k}$ and
$\vect{k}'$, it can be expanded in a Taylor series [treating $\vect{q}
= \vect{k} - \vect{k}'$ and $\vect{Q} = \frac12 (\vect{k} +
\vect{k}')$ as the independent variables], with the result
\begin{multline}
   \Lambda_{ss'} (\vect{k}, \vect{k}'; \vect{G}, \vect{G}'; \omega) =
   [ \delta_{ss'} \delta_{\vect{G} \vect{G}'} - \partial
   \Sigma^{(0)}_{ss'} (\vect{G}, \vect{G}'; \omega) / \partial \omega
   ] + q_{\alpha} \Lambda_{ss'}^{(\alpha|\cdot)} (\vect{G}, \vect{G}';
   \omega) + Q_{\alpha} \Lambda_{ss'}^{(\cdot|\alpha)} (\vect{G},
   \vect{G}'; \omega) \\ + q_{\alpha} q_{\beta}
   \Lambda_{ss'}^{(\alpha\beta|\cdot)} (\vect{G}, \vect{G}'; \omega) +
   Q_{\alpha} Q_{\beta} \Lambda_{ss'}^{(\cdot|\alpha\beta)} (\vect{G},
   \vect{G}'; \omega) + q_{\alpha} Q_{\beta}
   \Lambda_{ss'}^{(\alpha|\beta)} (\vect{G}, \vect{G}'; \omega) +
   \order (q^3) .  \label{eq:Lambda_Taylor}
\end{multline}
Here the lowest-order term is determined by the Ward identity
(\ref{eq:Ward_Gamma}), and the Taylor series coefficients such as
$\Lambda_{ss'}^{(\alpha|\beta)}$ are given in Appendix
\ref{app:Taylor}.  The analytic potential (\ref{eq:V1a}) can be
expanded in the same way:
\begin{multline}
   W^{(1)}_{ss'} (\vect{k}, \vect{k}'; \vect{G}, \vect{G}'; \omega) =
   W^{(1)}_{ss'} (\vect{0}, \vect{0}; \vect{G}, \vect{G}'; \omega) +
   q_{\alpha} W_{ss'}^{(\alpha|\cdot)} (\vect{G}, \vect{G}'; \omega) +
   Q_{\alpha} W_{ss'}^{(\cdot|\alpha)} (\vect{G}, \vect{G}'; \omega)
   \\ + q_{\alpha} q_{\beta} W_{ss'}^{(\alpha\beta|\cdot)} (\vect{G},
   \vect{G}'; \omega) + Q_{\alpha} Q_{\beta}
   W_{ss'}^{(\cdot|\alpha\beta)} (\vect{G}, \vect{G}'; \omega) +
   q_{\alpha} Q_{\beta} W_{ss'}^{(\alpha|\beta)} (\vect{G}, \vect{G}';
   \omega) + \order (q^3) .  \label{eq:V1_Taylor}
\end{multline}

In the expansion (\ref{eq:Lambda_Taylor}), a term such as $q_{\alpha}$
appears in the total potential (\ref{eq:V1}) as a multiplicative
factor in front of the screened potential $\varphi(\vect{q})$.  In
coordinate space, this term therefore takes the gradient of
$\varphi(\vect{x})$, generating the $\alpha$ component of the
macroscopic electric field produced by the perturbation $v(\vect{x})$.
Likewise, the term $Q_{\alpha}$ has the form of a (symmetrized)
crystal momentum operator that acts upon the envelope functions in an
effective-mass theory.  The Taylor series (\ref{eq:V1_Taylor}) for the
analytic potential (\ref{eq:V1a}) is interpreted in the same way,
except that these terms produce only short-range localized potentials
because they are analytic functions of $\vect{q}$.

The various terms in Eq.\ (\ref{eq:Lambda_Taylor}) therefore give rise
to various long-range spin-dependent potentials whose particular form
is determined by the symmetry of the coefficients
$\Lambda_{ss'}^{(\alpha|\cdot)}$, etc.  The specific term that
generates the long-range Rashba effect is
$\Lambda_{ss'}^{(\alpha|\beta)}$, since this term is linear in the
electric field $q_{\alpha} \varphi(\vect{q})$ and linear in the
crystal momentum $Q_{\beta}$.  The usual short-range part of the
Rashba coupling is generated by the analogous term
$W_{ss'}^{(\alpha|\beta)}$ in Eq.\ (\ref{eq:V1_Taylor}).

The complete expression for $V^{(1)}$ is obtained by inserting the
expansion (\ref{eq:Lambda_Taylor}) for $\Lambda$ and one of the three
expansions (\ref{eq:phi_q_exp_a}), (\ref{eq:phi_q_exp_b}), or
(\ref{eq:phi_q_exp_c}) for $\varphi(\vect{q})$ into Eq.\
(\ref{eq:V1}).  For the specific example of isovalent perturbations in
zinc-blende materials treated in Eq.\ (\ref{eq:multipole1}), one finds
\begin{multline}
   V^{(1)}_{ss'} (\vect{k} + \vect{G}, \vect{k}' + \vect{G}'; \omega)
   = \frac{1 - \delta_{\vect{q} \vect{0}}}{q^2} \{ [ \delta_{ss'}
   \delta_{\vect{G} \vect{G}'} - \partial \Sigma^{(0)}_{ss'}
   (\vect{G}, \vect{G}'; \omega) / \partial \omega ] [ 4 \pi \bar{n}_2
   q^2 / \epsilon + C_3 q_x q_y q_z + C_4 (q_x^4 + q_y^4 + q_z^4)] \\
   + C_3 q_x q_y q_z [ q_{\alpha} \Lambda_{ss'}^{(\alpha|\cdot)}
   (\vect{G}, \vect{G}'; \omega) + Q_{\alpha}
   \Lambda_{ss'}^{(\cdot|\alpha)} (\vect{G}, \vect{G}'; \omega) ] \} +
   \text{analytic terms} + \order (q^3) , \label{eq:V1ZB}
\end{multline}
where the analytic terms include $W^{(1)}$ and contributions from the
analytic part of $\varphi (\vect{q})$.  From this result it can be
seen that the Rashba effect in isovalent zinc-blende materials does
not include any long-range terms (to within the accuracy of the
present approximation scheme), since $\Lambda_{ss'}^{(\alpha|\beta)}$
contributes only to $\order (q^3)$.  However, there are other
long-range spin-splitting terms of $\order (q^2)$ or lower, and the
Rashba effect does contribute nonnegligible long-range terms for
perturbations in classes I and II.

\subsection{Quadratic terms}

Turning now to the quadratic response, the two contributions to
$V^{(2)}$ in Eq.\ (\ref{eq:V1V2_phi}) will be denoted
$V^{(2\text{a})}$ and $V^{(2\text{b})}$, respectively.  The first of
these is given by
\begin{equation}
   V^{(2\text{a})}_{ss'} (\vect{k} + \vect{G}, \vect{k}' + \vect{G}';
   \omega) = \frac{1}{2} {\sum_{\vect{k}_1}}' \sum_{\vect{G}_1}
   \sum_{\vect{G}_2} \tilde{\Gamma}^{(2)}_{ss'} (\vect{k} + \vect{G},
   \vect{k}' + \vect{G}'; \vect{k}_1 + \vect{G}_1, \vect{q} -
   \vect{k}_1 + \vect{G}_2; \omega) \varphi (\vect{k}_1 + \vect{G}_1)
   \varphi (\vect{q} - \vect{k}_1 + \vect{G}_2) . \label{eq:V2a}
\end{equation}
Upon inserting Eq.\ (\ref{eq:phi_qG_2}) for $\varphi (\vect{k} +
\vect{G})$ into the right-hand side, one obtains an expression for
$V^{(2\text{a})}$ very similar to that found in Eqs.\ (\ref{eq:N_ABC})
and (\ref{eq:ABC}) for the effective quadratic density
$\bar{n}^{(2)}(\vect{q})$.  Just as before, there are both analytic
and nonanalytic contributions.  The nonanalytic contributions can be
evaluated using the method outlined in Appendix \ref{app:nonanalytic};
the results show that the nonanalytic terms in $V^{(2\text{a})}$ are
$\order (q^{-1})$ for class I, $\order (q)$ for class II, and
$\order (q^3)$ for class III.  (An explicit expression for the
leading nonanalytic term in class I was given previously by
Sham. \cite{Sham66,Sak68}) Therefore, the nonanalytic contributions
are negligible in all three cases, and the leading term in
$V^{(2\text{a})}$ is just a constant:
\begin{equation}
   V^{(2\text{a})}_{ss'} (\vect{k} + \vect{G}, \vect{k}' + \vect{G}';
   \omega) = V^{(2\text{a})}_{ss'} (\vect{G}, \vect{G}'; \omega) +
   \order (q) . \label{eq:V2a_2}
\end{equation}
This term is negligible under the approximation schemes for classes I
and II.

Finally, the second contribution to $V^{(2)}$ in Eq.\
(\ref{eq:V1V2_phi}) is given by
\begin{equation}
   V^{(2\text{b})}_{ss'} (\vect{k} + \vect{G}, \vect{k}' + \vect{G}';
   \omega) = \sum_{\vect{G}''} \tilde{\Gamma}^{(1)}_{ss'} (\vect{k} +
   \vect{G}, \vect{k}' + \vect{G}'; \vect{q} + \vect{G}''; \omega)
   \varphi^{(2)} (\vect{q} + \vect{G}'') , \label{eq:V2b_1}
\end{equation}
where $\varphi^{(2)} (\vect{q} + \vect{G}'')$ was given previously in
Eq.\ (\ref{eq:phi2_qG}).  Inserting this expression into Eq.\
(\ref{eq:V2b_1}), one obtains
\begin{equation}
   V^{(2\text{b})}_{ss'} (\vect{k} + \vect{G}, \vect{k}' + \vect{G}';
   \omega) = \Lambda_{ss'} (\vect{k}, \vect{k}'; \vect{G}, \vect{G}';
   \omega) \varphi^{(2)} (\vect{q}) + W^{(2\text{b})}_{ss'}
   (\vect{k}, \vect{k}'; \vect{G}, \vect{G}'; \omega) , \label{eq:V2b}
\end{equation}
in which $\Lambda$ was defined in Eq.\ (\ref{eq:Lambda_ss}), and
\begin{equation}
   W^{(2\text{b})}_{ss'} (\vect{k}, \vect{k}'; \vect{G}, \vect{G}';
   \omega) = \sum_{\vect{G}'' \ne \vect{0}} \tilde{\Gamma}^{(1)}_{ss'}
   (\vect{k} + \vect{G}, \vect{k}' + \vect{G}'; \vect{q} + \vect{G}'';
   \omega) \xi^{(2)}_{\vect{G}''} (\vect{q}) . \label{eq:V2b1}
\end{equation}
Unlike the case for Eq.\ (\ref{eq:V1a}), this is not an analytic
function of $\vect{q}$.  However, as discussed below Eq.\
(\ref{eq:phi2_qG}), the nonanalytic portion is $\order (q)$ and
therefore vanishes at $q = 0$.  An explicit expression for
$V^{(2\text{b})}$ can now be obtained by inserting the expansion
(\ref{eq:phi2_q_N}) for $\varphi^{(2)} (\vect{q})$ into Eq.\
(\ref{eq:V2b}):
\begin{multline}
   V^{(2\text{b})}_{ss'} (\vect{k} + \vect{G}, \vect{k}' + \vect{G}';
   \omega) = q_{\alpha} w_c (\vect{q}) \{ \bar{n}^{(2)}_{\alpha} [
   \delta_{ss'} \delta_{\vect{G} \vect{G}'} - \partial
   \Sigma^{(0)}_{ss'} (\vect{G}, \vect{G}'; \omega) / \partial \omega
   + q_{\beta} \Lambda_{ss'}^{(\beta|\cdot)} (\vect{G}, \vect{G}';
   \omega) + Q_{\beta} \Lambda_{ss'}^{(\cdot|\beta)} (\vect{G},
   \vect{G}'; \omega) ] \} \\ + q_{\alpha} q_{\beta} w_c (\vect{q}) \{
   \bar{n}^{(2)}_{\alpha\beta} [ \delta_{ss'} \delta_{\vect{G}
   \vect{G}'} - \partial \Sigma^{(0)}_{ss'} (\vect{G}, \vect{G}';
   \omega) / \partial \omega ] \} + W^{(2\text{b})}_{ss'} (\vect{0},
   \vect{0}; \vect{G}, \vect{G}'; \omega) + \order (q) .
   \label{eq:V2b_2}
\end{multline}
The first and second terms are dipole and quadrupole potentials,
respectively, while the last term is just a constant.  In class II,
only the leading $\order (q^{-1})$ dipole term is retained; in
class I, the entire expression (\ref{eq:V2b_2}) is neglected.

\end{widetext}

\section{Nonlocal perturbations}

\label{sec:nonlocal}

An arbitrary nonlocal potential can be separated into local and
nonlocal parts (although this separation is not unique):
\begin{equation}
   v_{ss'} (\vect{x}, \vect{x}') = v^{\text{loc}}_{ss'} (\vect{x},
   \vect{x}') + v^{\text{nl}}_{ss'} (\vect{x}, \vect{x}') .
\end{equation}
Here the local part $v^{\text{loc}}_{ss'} (\vect{x}, \vect{x}')$ has
the form of Eq.\ (\ref{eq:v_local}) and is treated according to the
methods developed above.  This section considers the changes in the
preceding expressions that may be necessary in the case of nonlocal
perturbations, particularly those involving spin-orbit coupling.

\subsection{Analytic form}

A general nonlocal potential can be written as
\begin{equation}
   v_{ss'} (\vect{x}, \vect{x}') = \delta_{ss'} v_0 (\vect{x},
   \vect{x}') + \bm{\sigma}_{ss'} \cdot \vect{v} (\vect{x}, \vect{x}')
   ,
\end{equation}
where $v_0$ is a scalar relativistic potential, $\bm{\sigma}$ is the
Pauli matrix, and $\vect{v}$ is a pseudovector (similar to orbital
angular momentum) that accounts for spin-orbit coupling.  If $v$ is
hermitian and time-reversal invariant, then $v_0$ is real and
symmetric, while $\vect{v} (\vect{x}, \vect{x}') = - \vect{v}
(\vect{x}', \vect{x})$ is imaginary and antisymmetric.  Thus
$\vect{v}$ can have no local component, and a local
time-reversal-invariant potential must be a spin scalar:
\begin{equation}
   v^{\text{loc}}_{ss'} (\vect{x}, \vect{x}') = \delta_{ss'} \delta
   (\vect{x} - \vect{x}') v_{\text{loc}} (\vect{x}) .
\end{equation}

In the norm-conserving pseudopotential formalism, the nonlocal part of
the ionic pseudopotential $v^{\text{nl}}_{ss'} (\vect{x}, \vect{x}')$
is confined to a small region near the nucleus, typically either
having the form of a polynomial times a Gaussian
\cite{BaHaSc82,GoTeHu96,HaGoHu98} or vanishing absolutely outside a
core region of radius $r_c$. \cite{Kerk80,TroMar91}
As a result, $v^{\text{nl}}_{ss'} (\vect{k}, \vect{k}')$ is an entire
analytic function of $\vect{k}$ and $\vect{k}'$.


It is important to note that this analytic form relies upon a physical
approximation.  In an all-electron calculation where the
pseudopotential approximation is not used, the spin-orbit coupling
does in general include a contribution from the long-range Coulomb
part of the ionic potential.  The choice of an analytic
pseudopotential $v^{\text{nl}}_{ss'} (\vect{k}, \vect{k}')$ is
therefore an approximation, in which the spin-orbit coupling is
assumed to be dominated by the contribution from the ionic core.
Conventional norm-conserving pseudopotentials incorporate all
relativistic corrections of order $Z^2 \alpha^2$ (where $Z$ is the
atomic number and $\alpha$ is the fine-structure constant), but
neglect various terms of order $\alpha^2$, \cite{Klein80,BacSch82}
including the spin-orbit coupling from the long-range (but slowly
varying) Coulomb potential outside the core region.  This
approximation is used in all that follows.  It greatly simplifies the
analysis of spin-dependent perturbations, as shown below.

\subsection{Screening}

Consider now the description of screening for a spin-dependent
perturbation.  The relationship between the total polarization $\Pi$
and the regular polarization $P$ was given above in Eq.\
(\ref{eq:Dys_Pi_P}) for a general nonlocal potential.  Setting
$\vect{G} = \vect{0}$ in this equation gives
\begin{equation}
   \Pi_{ss'} (\vect{q}; \vect{k}, \vect{k} + \vect{q} + \vect{G}') =
   \epsilon^{-1} (\vect{q}) P_{ss'} (\vect{q}; \vect{k}, \vect{k} +
   \vect{q} + \vect{G}') , \label{eq:Pi_P_nl_1}
\end{equation}
where $\epsilon (\vect{q})$ is the same scalar dielectric function
defined above in Eq.\ (\ref{eq:eps_k}).  Substituting this result into
Eq.\ (\ref{eq:Dys_Pi_P}) then yields
\begin{multline}
   \Pi_{ss'} (\vect{q} + \vect{G}; \vect{k}, \vect{k} + \vect{q} +
   \vect{G}') = P_{ss'} (\vect{q} + \vect{G}; \vect{k}, \vect{k} +
   \vect{q} + \vect{G}') \\ + P(\vect{q} + \vect{G}, \vect{q})
   \frac{v_c(\vect{q})}{ \epsilon (\vect{q})} P_{ss'} (\vect{q}; \vect{k},
   \vect{k} + \vect{q} + \vect{G}') . \label{eq:Pi_P_nl_2}
\end{multline}
Equations (\ref{eq:Pi_P_nl_1}) and (\ref{eq:Pi_P_nl_2}) replace the
scalar equations (\ref{eq:Pi_P}) derived previously.

If the perturbation is now separated into local and nonlocal parts,
the linear screened potential (\ref{eq:phi_1a}) can be written in a
form similar to (\ref{eq:phi_nbar}):
\begin{equation}
   \varphi_{ss'} (\vect{k}, \vect{k}') = v^{\text{an}}_{ss'}
   (\vect{k}, \vect{k}') + \delta_{ss'} \frac{v_c(\vect{q}) [\bar{n}
   (\vect{q}) + n_{\text{nl}} (\vect{q})]}{\epsilon(\vect{q})} ,
   \label{eq:phi_1d}
\end{equation}
in which $v^{\text{an}}_{ss'} (\vect{k}, \vect{k}') =
v^{\text{nl}}_{ss'} (\vect{k}, \vect{k}') + \delta_{ss'} v_{\text{an}}
(\vect{q})$, $\vect{q} = \vect{k} - \vect{k}'$, $\bar{n} (\vect{q})$
is the effective density (\ref{eq:n_bar}) for the local potential, and
\begin{multline}
   n_{\text{nl}} (\vect{q}) = \sum_{\vect{k}''} \sum_{\vect{G}}
   P_{\lambda' \lambda} (\vect{q} ; \vect{k}'', \vect{k}'' + \vect{q}
   + \vect{G}) \\ \times v^{\text{nl}}_{\lambda \lambda'} (\vect{k}''
   + \vect{q} + \vect{G}, \vect{k}'') \label{eq:n_nl}
\end{multline}
is a correction to $\bar{n} (\vect{q})$ from the nonlocal part of the
perturbation.  Since $n_{\text{nl}} (\vect{q}) = \order (q)$ is
an analytic function with the same site symmetry as $\bar{n}
(\vect{q})$, this term does not produce any qualitative changes in
$\varphi$; it merely renormalizes $\bar{n} (\vect{q})$.  The only
qualitatively new contribution to $\varphi$ is the spin-dependent term
$v^{\text{nl}}_{ss'} (\vect{k}, \vect{k}')$ itself, which is analytic.

For wave vectors in the vicinity of a nonzero reciprocal lattice
vector, it is convenient to write Eq.\ (\ref{eq:phi_1a}) in the
following alternative form:
\begin{equation}
   \varphi_{ss'} (\vect{k} + \vect{G}, \vect{k}') =
   R_{\vect{G}\vect{0}} (\vect{q}) \varphi_{ss'} (\vect{k}, \vect{k}')
   + \xi_{ss'}^{\vect{G}} (\vect{k}, \vect{k}') .  \label{eq:phi_1e}
\end{equation}
Here $R_{\vect{G}\vect{0}} (\vect{q})$ was defined in Eq.\
(\ref{eq:RG0}), and
\begin{multline}
   \xi_{ss'}^{\vect{G}} (\vect{k}, \vect{k}') = \delta_{ss'} [
   \xi_{\vect{G}} (\vect{q}) + (1 - \delta_{\vect{G}\vect{0}})
   v_c(\vect{q} + \vect{G}) n_{\text{nl}} (\vect{q} + \vect{G}) ] \\ +
   v^{\text{nl}}_{ss'} (\vect{k} + \vect{G}, \vect{k}') -
   R_{\vect{G}\vect{0}} (\vect{q}) v^{\text{nl}}_{ss'} (\vect{k},
   \vect{k}')
\end{multline}
is a generalization of the function $\xi_{\vect{G}} (\vect{q})$
defined in Eq.\ (\ref{eq:chi_G}).  This is an analytic function of
$\vect{k}$ and $\vect{k}'$ for $q < G_{\text{min}}$.

If Eqs.\ (\ref{eq:phi_1d}) and (\ref{eq:phi_1e}) are inserted into the
nonlocal version of Eq.\ (\ref{eq:V1_phi}) [i.e., Eq.\
(\ref{eq:V1V2_phi})], it is apparent that the nonlocal part of the
perturbation produces no qualitative change in the total linear
potential $V^{(1)}$.  The only change is a simple renormalization of
the analytic and nonanalytic terms in $V^{(1)}$.

The same conclusion also holds for the quadratic potential $V^{(2)}$.
Thus, the correct qualitative form of $V^{(1)}$ and $V^{(2)}$ can be
derived by ignoring the nonlocal (and spin-dependent) part of the
perturbing potential, and including spin only in the vertex function
$\Lambda$ and the analytic parts of $V^{(1)}$ and $V^{(2)}$.  This is
precisely the approach used in Secs.\ \ref{sec:linear},
\ref{sec:quadratic}, and \ref{sec:selfenergy}.

The key to obtaining this simple result is the fact that
$v^{\text{nl}}_{ss'} (\vect{k}, \vect{k}')$ is analytic.  As shown
above, this relies upon the approximation of neglecting spin-orbit
coupling outside the atomic cores.  Such an approximation would also
be possible (and even desirable for its simplicity) in an all-electron
calculation where the core electrons are treated explicitly.

\section{Summary and conclusions}

\label{sec:conclusions}

This paper has presented an analysis of the self-energy of an electron
in a lattice-matched semiconductor heterostructure for small values of
the crystal momentum.  A general theory of nonlinear response for
nonlocal spin-dependent perturbations was developed in terms of vertex
functions and the static polarization, and applied to the case of
quadratic response in a periodic insulator at zero temperature.  A set
of Ward identities was established for nonlocal spin-dependent
potentials.  The heterostructure perturbation was separated into a
local spin-independent part and a nonlocal spin-dependent part, and
the contributions from these were analyzed separately.  Due to the
neglect of spin-orbit coupling outside the atomic cores, the nonlocal
part of the potential is analytic in momentum space.  As a result, the
nonlocal part of the perturbation merely renormalizes the
contributions from the local part.

The main results of the paper are presented in Eqs.\ (\ref{eq:V1}),
(\ref{eq:V2a_2}), and (\ref{eq:V2b}).  The total linear potential
(\ref{eq:V1}) has the form $V^{(1)} = \Lambda \varphi + W^{(1)}$, in
which all of the nonanalytic contributions come from the screened
scalar potential $\varphi$.  This has a form $\varphi(\vect{q}) =
v_{\text{an}} (\vect{q}) + v_c (\vect{q}) \bar{n} (\vect{q}) /
\epsilon (\vect{q})$ similar to that for screening in a homogeneous
system, except that the effective density $\bar{n} (\vect{q})$ has the
site symmetry of the perturbation and the macroscopic dielectric
function $\epsilon (\vect{q})$ has the symmetry of the reference
crystal.

Spin-dependent contributions come from the analytic part $W^{(1)}$ and
the vertex function $\Lambda$.  The vertex function can be expanded in
a Taylor series (\ref{eq:Lambda_Taylor}), in which $q_{\alpha}$ takes
the gradient (in coordinate space) of $\varphi(\vect{q})$, while
$Q_{\alpha}$ is a crystal momentum operator in effective-mass theory.
The generalized Rashba effect comes from the term linear in
$q_{\alpha}$ and $Q_{\beta}$, but there are other spin-splitting
contributions from the lower-order terms as well.  A more detailed
analysis of the various terms is given in the following paper on
effective-mass theory. \cite{Fore05b}

The total quadratic potential of Eqs.\ (\ref{eq:V2a_2}) and
(\ref{eq:V2b}) has a similar form $V^{(2)} = \Lambda \varphi^{(2)} +
W^{(2)}$, in which $W^{(2)}$ is analytic to within the accuracy of the
approximation scheme defined in the Introduction.  The quadratic
screened potential is $\varphi^{(2)} (\vect{q}) = v_c (\vect{q})
\bar{n}^{(2)} (\vect{q}) / \epsilon (\vect{q})$, where the effective
external density $\bar{n}^{(2)} (\vect{q})$ has the site symmetry of a
diatomic perturbation in the reference crystal.  Due to the Ward
identities for an insulator, the leading term in the power series
expansion of $\bar{n}^{(2)} (\vect{q})$ is a dipole term.  The Rashba
term in the quadratic potential is always negligible under the
approximation scheme used here.

The results derived in this paper are used to develop a
first-principles effective-mass theory in the following paper.
\cite{Fore05b} The present results are of crucial importance in
establishing clearly defined limitations on the validity of this
theory.  Most previous formulations of effective-mass theory have been
based on non-self-consistent empirical pseudopotentials, for which the
possibility of long-range Coulomb interactions is not even considered.
However, as shown here, long-range potentials arising from nonanalytic
terms in the screening potential---and even the charge density
itself---must be considered in general.

The omission of such terms is partially justified (to a certain order
of approximation) in isovalent zinc-blende systems, where high crystal
symmetry eliminates the contributions from dipole and quadrupole terms
in the linear response. \cite{BaReBaPe89} However, it is not fully
justified even in zinc-blende, since the leading octopole terms are of
a lower order than the position dependence of the effective mass,
which is often included in heterostructure effective-mass
calculations.  The following paper \cite{Fore05b} accounts for all
terms of the same order as the position dependence of the effective
mass, including the octopole and hexadecapole potentials derived here
in Eqs.\ (\ref{eq:multipole1}) and (\ref{eq:V1ZB}).

A pioneering paper by Sham on effective-mass theory for shallow
impurity states \cite{Sham66} dealt with many of the same issues (for
local spin-independent potentials), but at a lower order of
approximation.  In particular, Sham considered only the lowest-order
terms in cubic crystals.  At this level of approximation, the total
polarization can be treated as analytic [see Eq.\ (4.8) of Ref.\
\onlinecite{Sham66}], whereas the present Eq.\ (\ref{eq:Pi_P}) shows
that this is no longer true for terms of higher order (such as those
needed for the analysis in Ref.\ \onlinecite{Fore05b}) or crystals of
lower symmetry.  The present work provides a systematic framework for
extending Sham's analysis to crystals of general symmetry and terms of
arbitrary order.

The value of establishing such a framework is demonstrated by the
result (\ref{eq:phi2_q_N}) derived here for the leading dipole term in
the quadratic density response.  Sham has stated that the quadratic
density response contains no dipole terms, \cite{note:dipole} but the
justification for this statement is not clear because no details of
his calculation were given.  However, numerical evidence to the
contrary was subsequently provided by Dandrea, Duke, and Zunger in a
first-principles study of band offsets in InAs/GaSb
superlattices. \cite{DanZun92} They deduced that the calculated
difference between the macroscopic interface dipoles for InSb and GaAs
interfaces must be a nonlinear effect (because such differences do not
exist in linear response theory \cite{BaReBaPe89} in cubic crystals),
but did not inquire further into its origin.

To the author's knowledge, the present derivation provides the first
direct explanation for their result, and the first demonstration that
dipole terms are a general feature of the quadratic density response.
The magnitude of such dipoles is small---contributing 50 to 100 meV to
the band offset of typical no-common-atom
heterojunctions\cite{DanZun92,Seidel97}---but they play an important
role in explaining the experimentally observed asymmetry of band
offsets in such systems. \cite{Seidel97}

As a final note, it is worth drawing attention to a fundamental
property of the nonlinear response of insulators that apparently is
not widely known.  For example, in Refs.\ \onlinecite{BaReBaPe89} and
\onlinecite{PBBR90}, Baroni {\em et al}.\ have pointed out that
``within linear response theory, the electronic charge induced by a
given perturbation is proportional to the charge of the perturbing
potential,'' \cite{PBBR90} which implies that ``within linear response
theory, isovalent substitutional impurities carry no net charge.''
\cite{BaReBaPe89} Although the restriction to linear response theory
is necessary in general, the results derived here (in Sec.\
\ref{sec:quadratic}) demonstrate that the quadratic density response
of an {\em insulator} to a charged perturbation also carries no net
charge (i.e., it vanishes in the limit of small wave vectors).
Indeed, upon replacing Eqs.\ (\ref{eq:Ward_Pi}) and
(\ref{eq:Pi_scr_2}) with their higher-order generalizations, one finds
that this statement remains true for the nonlinear density response
(\ref{eq:n_nu}) of arbitrary order.

This result stems from the Ward identities (\ref{eq:Ward_Pi}) for the
total static polarization and proper polarization (see Sec.\
\ref{sec:screen_polarization}) of an insulator at zero temperature.
As a consequence of these identities, one can therefore state that in
an insulator, the total electronic charge induced by a given
perturbation is exactly linearly proportional to the charge of the
perturbing potential.  Of course, this statement assumes that the
system remains insulating over the full range of the perturbation
(from zero to full strength); otherwise, the perturbation theory used
here is no longer valid.

\begin{acknowledgments}
This work was supported by Hong Kong UGC Grant HIA03/04.SC02.
\end{acknowledgments}

\appendix

\section{Reducing heterovalent perturbations to isovalent perturbations}

\label{app:cluster}

In a heterovalent system such as Ge/GaAs, \cite{Har78} the ionic
perturbations are from class I.  However, it is often possible to
reduce such problems to an equivalent class II or III problem, because
accumulations of macroscopic charge are energetically unfavorable;
therefore, the interfaces grown in real heterojunctions tend to be
macroscopically neutral.

For Ge/GaAs, an ideal (110) heterojunction is already neutral, but an
ideal (001) or (111) interface would have a large macroscopic
interface charge, leading to a large compensating interface
free-carrier density. \cite{Har78} Since this is not observed
experimentally, the atoms in a real interface are believed to be
arranged in one or more layers of mixed composition, such that the net
macroscopic interface charge is zero.  \cite{BaReBaPe89,Har78} (This
is similar to the concept of surface reconstruction, but the interface
layers differ from the bulk only in chemical composition, not
structure.)  In such a system it is possible to replace the
heterovalent ionic perturbations with a set of equivalent isovalent
perturbations, simply by grouping the atoms together in clusters.

The first step is to define quasiatomic building blocks using a
modified version of Evjen's technique. \cite{Evjen32} Let $\Omega_0
(\vect{r})$ be a Wigner-Seitz cell (of the reference crystal) that is
centered on position $\vect{r}$, and let $\mathcal{N}$ be the number
of atoms in any primitive cell of the reference crystal.  For a given
atom $a$ at position $\vect{r}_a$ in the heterostructure, the
quasiatomic potential $\bar{v}_a$ is defined in terms of the ionic
potentials $v_{a'}$ for all atoms $a'$ via
\begin{equation}
   \bar{v}_a = \sum_{a'} w_{a'}(a) v_{a'} .
\end{equation}
Here $w_{a'}(a)$ is a weight factor, defined as $w_{a'}(a) =
1/\mathcal{N}$ if atom $a'$ lies inside $\Omega_0 (\vect{r}_a)$,
$w_{a'}(a) = 0$ if $a'$ lies outside $\Omega_0 (\vect{r}_a)$, and
$w_{a'}(a) = 1/m\mathcal{N}$ if $a'$ lies on the surface of $\Omega_0
(\vect{r}_a)$ [where $m$ is the number of cells $\Omega_0 (\vect{r}_a
+ \vect{R})$ that share atom $a'$, with $\vect{R}$ any Bravais lattice
vector of the reference crystal].  In a bulk crystal, these quasiatoms
are neutral objects with the site symmetry of atom $a$ in the
reference crystal.  Therefore, in a heterostructure, the quasiatoms
carry a charge only near the heterojunctions.

For Ge/GaAs, each quasiatomic building block is constructed from
$\frac12$ of the potential for a given ion plus $\frac18$ of the
potential for each of its four nearest neighbors.  In a bulk
zinc-blende crystal, these quasiatoms have $T_d$ symmetry and possess
no charge, no dipole moment, and no quadrupole moment.  Therefore, in
a heterostructure, the quasiatoms have monopole, dipole, and
quadrupole moments only near the heterojunctions.

It is assumed here that the ions in the mixed-composition interface
layers form a periodic array, so that a two-dimensional superlattice
translation symmetry exists in the directions parallel to the junction
plane.  In this case, one can define a three-dimensional
``slab-adapted'' \cite{Heine63,Klein81} unit cell of {\em quasiatoms}
that is large enough to contain 100\% of the ions in the
mixed-composition layers.  This unit cell has no net charge (since the
interface is assumed to be macroscopically neutral), and for some
choices of compositional mixing, it may also have no dipole
moment. \cite{Har78}

Thus, if one treats these slab-adapted unit cells as the fundamental
perturbations, this type of heterovalent class I perturbation can be
replaced by an equivalent neutral perturbation from class II or class
III.  In general the interface cells do have a dipole moment, so the
interface perturbations are class II, while the bulk perturbations are
class III.  However, an interface dipole term of $\order (q^{-1})$ is
physically equivalent to a bulk quadrupole term of $\order (q^0)$.
Therefore, the approximation scheme defined in the Introduction yields
results of the same overall accuracy for both the class II interface
and class III bulk perturbations in Ge/GaAs.

\section{Symmetry properties}

\label{app:symmetry}

This appendix considers some symmetry properties of $G$ and $\Sigma$.
Time-reversal symmetry is developed from the properties of zero- and
one-particle states.  The vacuum state $|0\rangle$ is defined to be
time-reversal invariant:
\begin{equation}
   \hat{\Theta} |0\rangle = |0\rangle , \label{eq:vacuum}
\end{equation}
where $\hat{\Theta}$ is the antiunitary time-reversal operator.  The
phase of $\hat{\Theta}$ may be partially defined by letting the
single-particle basis states $|\vect{x}, s \rangle =
\hat{\psi}_s^{\dag}(\vect{x}) |0\rangle$ satisfy
\begin{equation}
   \hat{\Theta} | \vect{x}, s \rangle = (-1)^{s - 1/2} | \vect{x}, -s
   \rangle .  \label{eq:tr_sp}
\end{equation}
This relation is consistent with the operator equation
\cite{Merz98_p612}
\begin{equation}
   \hat{\Theta} \hat{\psi}_s^{\dag}(\vect{x}) \hat{\Theta}^{\dag} =
   (-1)^{s - 1/2} \hat{\psi}_{-s}^{\dag}(\vect{x})
   . \label{eq:theta_def}
\end{equation}
One may therefore define $\hat{\Theta}$ over the entire many-particle Fock
space by Eqs.\ (\ref{eq:vacuum}) and (\ref{eq:theta_def}).

The next step is to use the identity \cite{Sak94_p273}
\begin{equation}
   \langle \beta | \hat{A} | \alpha \rangle = \langle \tilde{\beta} |
   \hat{\Theta} \hat{A} \hat{\Theta}^{\dag} | \tilde{\alpha} \rangle^*
   = \langle \tilde{\alpha} | \hat{\Theta} \hat{A}^{\dag}
   \hat{\Theta}^{\dag} | \tilde{\beta} \rangle , \label{eq:TRa}
\end{equation}
in which $\hat{A}$ is a linear operator and $|\tilde{\alpha}\rangle =
\hat{\Theta} | \alpha \rangle$.  If the many-particle Hamiltonian
$\hat{H}$ is time-reversal invariant (i.e., $[\hat{\Theta}, \hat{H}] =
0$), one has
\begin{equation}
   \hat{\Theta} [\hat{\psi}_s (\vect{x}, \tau) ]^{\dag}
   \hat{\Theta}^{\dag} = (-1)^{s - 1/2} \hat{\psi}_{-s}^{\dag}
   (\vect{x}, -\tau) ,
\end{equation}
which holds for complex $\tau$.  From this and Eq.\ (\ref{eq:TRa}) one
immediately obtains
\begin{equation}
   G_{ss'}(\vect{x}, \tau; \vect{x}', \tau') = (-1)^{s-s'}
   G_{-s',-s}(\vect{x}', -\tau'; \vect{x}, -\tau) ,
   \label{eq:reciprocity}
\end{equation}
which is the generalization of an ordinary Green-function
``reciprocity'' relation \cite{MorsFesh53_recip} to the
interacting-particle case.  (A similar expression was given in Ref.\
\onlinecite{Noz64}, but with the sign term omitted.) Since the change
of time variables in (\ref{eq:reciprocity}) does not alter $\tau -
\tau'$, the Fourier transform of (\ref{eq:reciprocity}) is just
\begin{equation}
   G_{ss'}(\vect{x}, \vect{x}', \omega) = (-1)^{s-s'}
   G_{-s',-s}(\vect{x}', \vect{x}, \omega) . \label{eq:G_tr}
\end{equation}
Now $\hat{H}$ is time-reversal invariant if and only if $h$ is, which
from (\ref{eq:tr_sp}) and (\ref{eq:TRa}) implies that
\begin{equation}
   h_{ss'}(\vect{x}, \vect{x}') = (-1)^{s-s'}
   h_{-s',-s}(\vect{x}', \vect{x}) .
\end{equation}
Equation (\ref{eq:Sigma_def}) then shows that $\Sigma$ has the same
time-reversal properties as $G$:
\begin{equation}
   \Sigma_{ss'}(\vect{x}, \vect{x}', \omega) = (-1)^{s-s'}
   \Sigma_{-s',-s}(\vect{x}', \vect{x}, \omega) . \label{eq:Sigma_tr}
\end{equation}

$G$ and $\Sigma$ may also satisfy other conditions derived from linear
symmetries of $\hat{H}$.  Consider the linear many-particle operator
$\hat{Q}$ defined by
\begin{equation}
   \hat{Q} = \sum_{s,s'} \iint \hat{\psi}_s^{\dag}(\vect{x})
   q_{ss'}(\vect{x}, \vect{x}') \hat{\psi}_{s'} (\vect{x}') d^3 \! x
   \, d^3 \! x' ,  \label{eq:Q_def}
\end{equation}
in which $q$ is a linear single-particle operator.  $\hat{Q}$ obeys
the commutation relations
\begin{subequations}
\label{eq:Q_psi}
\begin{align}
   [ \hat{\psi}_s (\vect{x}), \hat{Q} ] & = \sum_{s'} \int
   q_{ss'}(\vect{x}, \vect{x}') \hat{\psi}_{s'} (\vect{x}') d^3 \! x'
   , \\ [ \hat{\psi}_s^{\dag} (\vect{x}), \hat{Q} ] & = -\sum_{s'}
   \int \hat{\psi}_{s'}^{\dag} (\vect{x}') q_{s's}(\vect{x}',
   \vect{x}) d^3 \! x' .
\end{align}
\end{subequations}
The commutator of any two such operators is another operator with the
same form:
\begin{equation}
   [\hat{Q}_1, \hat{Q}_2] = \hat{Q}_3 ; \qquad q_3 \equiv [q_1,q_2]
   . \label{eq:Q1Q2Q3}
\end{equation}
Now suppose that the Hamiltonian has the symmetry $[\hat{K}, \hat{Q}]
= [\hat{H}, \hat{Q}] = 0$.  From (\ref{eq:Q1Q2Q3}), this is possible
only if $[h, q] = 0$.  One can then use Eqs.\ (\ref{eq:G_def}) and
(\ref{eq:Q_psi}) and the cyclic property of the trace to show that
\begin{equation}
   [G, q] = 0 ,
\end{equation}
which further implies that $[\Sigma, q] = 0$.

This result is applied to lattice translations throughout this paper,
and to other space group operations in the following
paper. \cite{Fore05b} Also of interest in the present paper is the
spin operator $\vect{s}_{ss'}(\vect{x}, \vect{x}') = \frac{1}{2}
\bm{\sigma}_{ss'} \delta(\vect{x} - \vect{x}')$, where $\bm{\sigma}$
is the Pauli spin matrix vector.  If $h$ is independent of spin (i.e.,
$[h, \vect{s}] = 0$), then $[G, \vect{s}] = 0$, and $G$ and $\Sigma$
have the scalar form
\begin{equation}
   \Sigma_{ss'}(\vect{x}, \vect{x}', \omega) = \delta_{ss'}
   \Sigma(\vect{x}, \vect{x}', \omega) .
\end{equation}
However, if $h$ includes spin-orbit coupling (which is the case
studied here), $\Sigma$ is nondiagonal.

\section{Fourier transforms}

\label{app:Fourier}


The Fourier transforms of the potential with respect to momentum and
frequency are defined by
\begin{gather}
   V(\vect{k}, \vect{k}') = \frac{1}{\Omega} \int_{\Omega}
   \int_{\Omega} e^{-i \vect{k} \cdot \vect{x}} V(\vect{x}, \vect{x}')
   e^{i \vect{k}' \cdot \vect{x}'} d^3 \! x \, d^3 \!  x' ,
   \label{eq:fourier_V} \\ V(\zeta_n, \zeta_{n'}) = \frac{1}{\beta}
   \int_{0}^{\beta} \int_{0}^{\beta} e^{i \zeta_n \tau}
   V(\tau, \tau') e^{-i \zeta_{n'} \tau'} d\tau \, d\tau' .
\end{gather}
Since $V(\tau, \tau') = V(\tau - \tau')$, the latter integral is
always diagonal in $n$:
\begin{equation}
   V(\zeta_n, \zeta_{n'}) = V(\zeta_n) \delta_{nn'} .
\end{equation}
For many-variable quantities such as the vertex function
$\Gamma^{(\nu)}$, the Fourier integrals for $\Gamma^{(\nu)} (\vect{k},
\vect{k}'; \vect{q}, \vect{q}'; \ldots)$ have the same form as
(\ref{eq:fourier_V}) for each pair of $(\vect{k}, \vect{k}')$
variables.

For a function of the form $f(\vect{x}) = f(r) Y_l^m
(\hat{\vect{x}})$, where $Y_l^m$ is a spherical harmonic, the Fourier
transform is of the form $f(\vect{k}) = f(k) Y_l^m (\hat{\vect{k}})$,
where for $k > 0$,
\begin{equation}
   f(k) = \frac{4 \pi}{\Omega} (-i)^l \int_{0}^{\infty} r^2 f(r) j_l
   (kr) dr ,
\end{equation}
in which $j_l(kr)$ is a spherical Bessel function.  For the special
case $f(r) = r^{-n}$, Eq.\ (6.561.14) of Ref.\ \onlinecite{GraRyz94}
gives
\begin{equation}
   f(k) = \frac{4 \pi}{\Omega} (-i)^l \frac{\sqrt{\pi} \, \Gamma
   [\frac12 (l - n + 3)]}{2^{n-1} \Gamma [\frac12 (l + n)]} k^{n-3} ,
   \label{eq:fk}
\end{equation}
in which $\Gamma(z) = \int_{0}^{\infty} e^{-t} t^{z-1} dt$.  Equation
(\ref{eq:fk}) is valid for $k > 0$, $n > 1$, and $l > n -
3$. \cite{GraRyz94}

\section{Perturbation theory}

\label{app:perturbation}

The starting point for the perturbation theory used in Sec.\
\ref{sec:density} is the standard formula \cite{AbrGorDzy75,FetWal03}
\begin{equation}
   \langle \hat{A}_H(\tau) \rangle = \frac{\langle T_{\tau}
   [\hat{A}_I(\tau) \hat{\mathcal{U}} ] \rangle_0}{\langle
   \hat{\mathcal{U}} \rangle_0} ,
\end{equation}
where $\hat{A}_H(\tau)$ is a Heisenberg picture operator,
$\hat{A}_I(\tau)$ is the same operator in the interaction picture,
$\langle \hat{O} \rangle$ denotes the thermal average
(\ref{eq:thermal_ave}) with respect to $\hat{K}$, $\langle \hat{O}
\rangle_0$ is a thermal average with respect to $\hat{K}_0 = \hat{H}_0
- \mu \hat{N}$, and $\hat{\mathcal{U}} = T_{\tau} \{ \exp [
-\int_{0}^{\beta} \hat{H}_1(\tau) d\tau ] \}$.  If $\hat{\mathcal{U}}$
is expanded in a power series, terms of equal order in the numerator
and denominator can be grouped together as
\begin{multline}
   \langle \hat{A}_H(\tau) \rangle = \langle \hat{A}_I(\tau) \rangle_0
   + \sum_{\nu = 1}^{\infty} \frac{(-1)^{\nu}}{\nu !} \int_0^{\beta}
   d\tau_1 \cdots \int_0^{\beta} d\tau_{\nu} \\ \times \langle T_\tau
   [ \Delta \hat{H}_1 (\tau_1) \cdots \Delta \hat{H}_1 (\tau_{\nu})
   \Delta \hat{A}_I(\tau) ] \rangle_0 , \label{eq:pert}
\end{multline}
where $\Delta \hat{A}_{I,H}(\tau) = \hat{A}_{I,H} (\tau) - \langle
\hat{A}_I (\tau) \rangle_0$.  A more compact expression for
(\ref{eq:pert}) is
\begin{equation}
   \langle \Delta \hat{A}_H(\tau) \rangle = \langle T_{\tau} [\Delta
   \hat{A}_I(\tau) \hat{\mathcal{W}} ] \rangle_0 ,
\end{equation}
where $\hat{\mathcal{W}} = T_{\tau} \{ \exp [ -\int_{0}^{\beta} \Delta
\hat{H}_1(\tau) d\tau ] \}$.

\section{Polarization}

\label{app:polarization}

The static polarization (\ref{eq:n_nu}) is defined by
\begin{multline}
   \Pi^{(\nu)} (00', 11', \ldots, \nu\nu') \\ = \int_0^{\beta} d\tau_1
   \cdots \int_0^{\beta} d\tau_{\nu} D^{(\nu)} (00', 11', \ldots,
   \nu\nu') , \label{eq:Pi_def}
\end{multline}
where $D$ is the dynamic polarization
\begin{multline}
  D^{(\nu)} (00', 11', \ldots, \nu\nu') \\ = (-1)^{\nu} \langle
  T_{\tau} [ \Delta \hat{\rho} (00') \Delta \hat{\rho} (11') \cdots
  \Delta \hat{\rho} (\nu\nu') ] \rangle_0 . \label{eq:D}
\end{multline}
Here a superfluous second time variable has been added (for notational
convenience) to the interaction picture operators according to the
definition
\begin{equation}
   \hat{\rho}_{ss'}(\vect{x}, \tau; \vect{x}', \tau') \equiv
   \hat{\rho}_{ss'}(\vect{x}, \vect{x}', \tau - \tau') ,
\end{equation}
where $\tau' \equiv 0$.  Since $D^{(\nu)}$ is periodic (with period
$\beta$) in all time variables $\tau_{\lambda}$, but depends on time
only via the intervals $\tau_{\lambda} - \tau_0$ (for $\lambda = 1, 2,
\ldots, \nu$), it follows that $\Pi^{(\nu)}$ is independent of time.

By definition, $D$ is symmetric with respect to interchange of any
pair of operators $\Delta \hat{\rho}$; thus
\begin{multline}
   D^{(\nu)} (\ldots, ii', \ldots, jj', \ldots) \\ = D^{(\nu)}
   (\ldots, jj', \ldots, ii', \ldots) . \label{eq:D_symm}
\end{multline}
Another constraint on $D$ can be derived from time-reversal symmetry
using the methods of Appendix \ref{app:symmetry}:
\begin{equation}
   D^{(\nu)} (00', 11', \ldots, \nu\nu') = (-1)^s D^{(\nu)}
   (\bar{0}'\bar{0}, \bar{1}'\bar{1}, \ldots, \bar{\nu}'\bar{\nu}) .
   \label{eq:D_tr}
\end{equation}
Here $(\bar{\lambda}) = (\vect{x}_{\lambda}, -s_{\lambda},
\tau_{\lambda})$ and
\begin{equation}
   s = \sum_{\lambda = 0}^{\nu} (s_{\lambda} - s_{\lambda}') .
\end{equation}
These symmetries are valid for the static polarization $\Pi$ as well
(with the time variables omitted).

\begin{widetext}

\section{Functions $A$, $B$, and $C$}

\label{app:ABC}

The functions $A$, $B$, and $C$ introduced in Eq.\ (\ref{eq:N_ABC})
are defined by
\begin{equation} \label{eq:ABC}
   \begin{split}
   A(\vect{k}, \vect{k}_1, \vect{k}_2) & = \sum_{\vect{G}}
   \sum_{\vect{G}_1} \sum_{\vect{G}_2} R_{\vect{0}\vect{G}} (\vect{k})
   \tilde{\Pi}^{(2)} (\vect{k} + \vect{G}, \vect{k}_1 + \vect{G}_1,
   \vect{k}_2 + \vect{G}_2) R_{\vect{G}_1 \vect{0}} (\vect{k}_1)
   R_{\vect{G}_2 \vect{0}} (\vect{k}_2) , \\
   B(\vect{k}, \vect{k}_1, \vect{k}_2) & = \sum_{\vect{G}}
   \sum_{\vect{G}_1} \sum_{\vect{G}_2} R_{\vect{0}\vect{G}} (\vect{k})
   \tilde{\Pi}^{(2)} (\vect{k} + \vect{G}, \vect{k}_1 + \vect{G}_1,
   \vect{k}_2 + \vect{G}_2) \xi_{\vect{G}_1} (\vect{k}_1)
   R_{\vect{G}_2 \vect{0}} (\vect{k}_2) , \\
   C(\vect{k}, \vect{k}_1, \vect{k}_2) & = \sum_{\vect{G}}
   \sum_{\vect{G}_1} \sum_{\vect{G}_2} R_{\vect{0}\vect{G}} (\vect{k})
   \tilde{\Pi}^{(2)} (\vect{k} + \vect{G}, \vect{k}_1 + \vect{G}_1,
   \vect{k}_2 + \vect{G}_2) \xi_{\vect{G}_1} (\vect{k}_1)
   \xi_{\vect{G}_2} (\vect{k}_2) .
   \end{split}
\end{equation}
\end{widetext}

\section{Nonanalytic terms in $\bar{n}^{(2)}_A (\vect{q})$}

\label{app:nonanalytic}

To leading order, the term $\bar{n}^{(2)}_A (\vect{q})$ in Eq.\
(\ref{eq:N_ABC}) is
\begin{equation}
   \bar{n}^{(2)}_A (\vect{q}) = \frac12 A_{\alpha \beta \gamma}
   q_{\alpha} {\sum_{\vect{k}}}' k_{\beta} (q_{\gamma} - k_{\gamma})
   \varphi (\vect{k}) \varphi (\vect{q} - \vect{k}) + \order (q^3) ,
   \label{eq:N_A}
\end{equation}
in which $A_{\alpha\beta\gamma}$ is the Taylor series coefficient
(\ref{eq:Pi_Taylor_2}).  The only contribution to Eq.\ (\ref{eq:N_A})
that is of order $q^2$ comes from the monopole terms in $\varphi
(\vect{q})$.  For any perturbation comprising a finite number of
atoms, the expansion (\ref{eq:phi_q_exp}) begins as $\varphi
(\vect{q}) = -Z_v w_c (\vect{q}) / \Omega + \order (q^{-1})$, where
$Z_v$ is the net valence charge of the ionic perturbation.

If this lowest-order term is considered, the value of Eq.\
(\ref{eq:N_A}) can be estimated by extending the summation to all
values of $\vect{k}$.  For a cubic crystal with scalar $\epsilon$, the
convolution can be performed using the Fourier transforms in Appendix
\ref{app:Fourier}; the result is
\begin{equation}
   \bar{n}^{(2)}_A (\vect{q}) = \frac{\pi^2 Z_v^2}{8 \Omega
   \epsilon^2} A_{\alpha \beta \gamma} \frac{q_{\alpha} q_{\beta}
   q_{\gamma}}{q} + \order (q^3) .  \label{eq:N_A_cubic}
\end{equation}
This shows explicitly that the leading nonanalytic term in
$\bar{n}^{(2)}_A (\vect{q})$ is $\order (q^2)$.

However, this term exists only for the heterovalent perturbations of
class I in the Introduction.  For the isovalent perturbations ($Z_v =
0$) of classes II and III, there is no $\order (q^2)$ term in
$\bar{n}^{(2)}_A (\vect{q})$.  A similar analysis shows that for class
II, $\bar{n}^{(2)}_A (\vect{q}) = \order (q^4)$, while for class III,
$\bar{n}^{(2)}_A (\vect{q}) = \order (q^6)$.  Therefore, according to
the approximation scheme adopted in this paper, $\bar{n}^{(2)}_A
(\vect{q})$ is negligible in all three cases.

\begin{widetext}

\section{Vertex function Taylor series}

\label{app:Taylor}

Since the proper vertex function $\tilde{\Gamma}^{(1)}$ is an analytic
function of $\vect{k}$ and $\vect{k}'$, it can be expanded in a Taylor
series in the variables $\vect{q} = \vect{k} - \vect{k}'$ and
$\vect{Q} = \frac12 (\vect{k} + \vect{k}')$.  This yields an
expression similar to Eq.\ (\ref{eq:Lambda_Taylor}):
\begin{multline}
   \tilde{\Gamma}^{(1)}_{ss'} (\vect{k} + \vect{G}, \vect{k}' +
   \vect{G}'; \vect{q} + \vect{G}''; \omega) =
   \tilde{\Gamma}^{(1)}_{ss'} (\vect{G}, \vect{G}'; \vect{G}'';
   \omega) + q_{\alpha} \tilde{\Gamma}^{(\alpha|\cdot)}_{ss'}
   (\vect{G}, \vect{G}'; \vect{G}''; \omega) + Q_{\alpha}
   \tilde{\Gamma}^{(\cdot|\alpha)}_{ss'} (\vect{G}, \vect{G}';
   \vect{G}''; \omega) \\ + q_{\alpha} q_{\beta}
   \tilde{\Gamma}^{(\alpha\beta|\cdot)}_{ss'} (\vect{G}, \vect{G}';
   \vect{G}''; \omega) + Q_{\alpha} Q_{\beta}
   \tilde{\Gamma}^{(\cdot|\alpha\beta)}_{ss'} (\vect{G}, \vect{G}';
   \vect{G}''; \omega) + q_{\alpha} Q_{\beta}
   \tilde{\Gamma}^{(\alpha|\beta)}_{ss'} (\vect{G}, \vect{G}';
   \vect{G}''; \omega) + \order (q^3) . \label{eq:Gamma_Taylor}
\end{multline}
The Taylor series coefficients in Eq.\ (\ref{eq:Lambda_Taylor}) for
the effective vertex function $\Lambda$ are therefore given by
\begin{equation}
   \begin{split}
   \Lambda_{ss'}^{(\alpha|\cdot)} (\vect{G}, \vect{G}'; \omega) & =
   \tilde{\Gamma}^{(\alpha|\cdot)}_{ss'} (\vect{G}, \vect{G}';
   \vect{0}; \omega) + \sum_{\vect{G}'' \ne \vect{0}}
   \tilde{\Gamma}^{(1)}_{ss'} (\vect{G}, \vect{G}'; \vect{G}'';
   \omega) R^{\alpha}_{\vect{G}''\vect{0}} , \\
   \Lambda_{ss'}^{(\cdot|\alpha)} (\vect{G}, \vect{G}'; \omega) & =
   \tilde{\Gamma}^{(\cdot|\alpha)}_{ss'} (\vect{G}, \vect{G}';
   \vect{0}; \omega) , \\
   \Lambda_{ss'}^{(\alpha\beta|\cdot)} (\vect{G}, \vect{G}'; \omega) & =
   \tilde{\Gamma}^{(\alpha\beta|\cdot)}_{ss'} (\vect{G}, \vect{G}';
   \vect{0}; \omega) + \sum_{\vect{G}'' \ne \vect{0}} [
   \tilde{\Gamma}^{(\alpha|\cdot)}_{ss'} (\vect{G}, \vect{G}';
   \vect{G}''; \omega) R^{\beta}_{\vect{G}''\vect{0}} +
   \tilde{\Gamma}^{(1)}_{ss'}
   (\vect{G}, \vect{G}'; \vect{G}''; \omega)
   R^{\alpha\beta}_{\vect{G}''\vect{0}} ] , \\
   \Lambda_{ss'}^{(\cdot|\alpha\beta)} (\vect{G}, \vect{G}'; \omega) & =
   \tilde{\Gamma}^{(\cdot|\alpha\beta)}_{ss'} (\vect{G}, \vect{G}';
   \vect{0}; \omega) , \\
   \Lambda_{ss'}^{(\alpha|\beta)} (\vect{G}, \vect{G}'; \omega) & =
   \tilde{\Gamma}^{(\alpha|\beta)}_{ss'} (\vect{G}, \vect{G}';
   \vect{0}; \omega) + \sum_{\vect{G}'' \ne \vect{0}}
   \tilde{\Gamma}^{(\cdot|\beta)}_{ss'} (\vect{G}, \vect{G}';
   \vect{G}''; \omega) R^{\alpha}_{\vect{G}''\vect{0}} ,
   \end{split}
\end{equation}
in which $R^{\alpha}_{\vect{G}\vect{0}}$ and
$R^{\alpha\beta}_{\vect{G}\vect{0}}$ are the Taylor series
coefficients for $R_{\vect{G}\vect{0}} (\vect{q})$.  In these
expressions, some special values of the coefficients for the case
$\vect{G}'' = \vect{0}$ are given by the Ward identity
(\ref{eq:Ward_Gamma}):
\begin{equation}
   \begin{split}
   \tilde{\Gamma}^{(1)}_{ss'} (\vect{G}, \vect{G}'; \vect{0}; \omega)
   & = \delta_{ss'} \delta_{\vect{G} \vect{G}'} - \frac{\partial
   \Sigma^{(0)}_{ss'} (\vect{G}, \vect{G}'; \omega)}{\partial \omega}
   , \\
   \tilde{\Gamma}^{(\cdot|\alpha)}_{ss'} (\vect{G}, \vect{G}';
   \vect{0}; \omega) & = - \frac{\partial}{\partial k_{\alpha}}
   \left. \frac{\partial \Sigma^{(0)}_{ss'} (\vect{k} + \vect{G},
   \vect{k} + \vect{G}'; \omega)}{\partial \omega} \right|_{\vect{k} =
   \vect{0}} , \\
   \tilde{\Gamma}^{(\cdot|\alpha\beta)}_{ss'} (\vect{G}, \vect{G}';
   \vect{0}; \omega) & = - \frac12 \frac{\partial^2}{\partial
   k_{\alpha} \partial k_{\beta}} \left. \frac{\partial
   \Sigma^{(0)}_{ss'} (\vect{k} + \vect{G}, \vect{k} + \vect{G}';
   \omega)}{\partial \omega} \right|_{\vect{k} = \vect{0}} .
   \end{split}
\end{equation}

\end{widetext}


\begin{thebibliography}{109}
\bibitem[{not({\natexlab{a}})}]{note:screening}
\bibinfo{note}{One could also write $\varphi (\vect{q}) = [v_{\text{an}}
  (\vect{q}) + v_c (\vect{q}) \bar{n} (\vect{q})] / \epsilon (\vect{q})$ if the
  second term in Eq.\ (\ref{eq:n_bar}) is omitted, but it is more convenient to
  let the term that is not proportional to $v_c (\vect{q})$ be an analytic
  function of $\vect{q}$.}
\bibitem[{not({\natexlab{b}})}]{note:class1}
\bibinfo{note}{If one wishes to go beyond the stated level of approximation,
  note that Eq.\ (\ref{eq:phi2_q_N}) is incomplete for class I perturbations,
  since the nonanalytic part of $\bar{n}^{(2)}_A (\vect{q})$ is $\order
  (q^2)$.}
\bibitem[{not({\natexlab{c}})}]{note:dipole}
\bibinfo{note}{More precisely, on p.\ 725 of Ref.\ \onlinecite{Sham66}, Sham
  says that the contribution from the quadratic density response to the
  self-energy is ``short-ranged,'' meaning that it contains no terms with a
  $1/r$ or $1/r^2$ dependence at large $r$.}
\end{thebibliography}

\end{document}